%% file: manuscript.tex
\documentclass[fleqn,usenatbib]{mnras}

\usepackage{newtxtext,newtxmath}

\usepackage[T1]{fontenc}
\usepackage{tabularx}

\DeclareRobustCommand{\VAN}[3]{#2}
\let\VANthebibliography\thebibliography
\def\thebibliography{\DeclareRobustCommand{\VAN}[3]{##3}\VANthebibliography}

\usepackage{graphicx}	
\usepackage{amsmath}	
\usepackage{orcidlink}
\usepackage{xcolor}
\usepackage[percent]{overpic}

\definecolor{red}{rgb}{0.96, 0.36, 0.36}

\title[Quantifying the MW and LMC with SBI]{Quantifying the Milky Way, LMC and their interaction using all-sky kinematics of outer halo stars}
\author[R. A. N. Brooks et al.]{Richard A. N. Brooks$^{1}$\thanks{E-mail: richard.brooks.22@ucl.ac.uk}\orcidlink{0000-0001-5550-2057},
Jason L. Sanders$^{1}$\orcidlink{0000-0003-4593-6788},
Adam M. Dillamore$^{1}$\orcidlink{0000-0003-0807-5261},
Nicolás Garavito-Camargo$^{2}$\thanks{NASA Hubble Fellowship Program, Einstein Fellow}\orcidlink{0000-0001-7107-1744}, \newauthor
Vedant Chandra$^{3}$\orcidlink{0000-0002-0572-8012},
Adrian~M.~Price-Whelan$^{4}$\orcidlink{0000-0003-0872-7098}, 
Phillip Cargile$^{3}$\orcidlink{0000-0002-1617-8917}
\\
$^{1}$Department of Physics and Astronomy, University College London, London, WC1E 6BT, UK\\
$^{2}$ Department of Astronomy, University of Maryland, College Park, MD 20742, USA\\
$^{3}$ Center for Astrophysics | Harvard \& Smithsonian, 60 Garden Street, Cambridge, MA 02138, USA\\
$^{4}$Center for Computational Astrophysics, Flatiron Institute, Simons Foundation, 162 Fifth Avenue, New York, NY 10010, USA\\
}

\date{Accepted XXX. Received YYY; in original form ZZZ}

\pubyear{\the\year{}}

\begin{document}
\label{firstpage}
\pagerange{\pageref{firstpage}--\pageref{lastpage}}
\maketitle

\begin{abstract}
The recent pericentric passage of the Large Magellanic Cloud (LMC) has dislodged the Milky Way's (MW) centre of mass, inducing dynamical disequilibrium, the \textit{reflex motion}, in the kinematics of outer stellar halo stars.
Using data out to $160 \, \rm kpc$ from the combined H3+SEGUE+MagE outer halo survey, we constrain the mass of the MW and LMC, as well as the resulting reflex motion and the stellar halo velocity anisotropy.
Using a suite of 32,000 rigid MW--LMC simulations, each with a MW stellar halo evolved to the present day in the combined MW--LMC potential, we perform Simulation Based Inference by training a neural posterior estimator on the means and dispersions of the radial and tangential velocities of stars from the combined H3+SEGUE+MagE outer halo sample.
Relative to halo stars at $100 \, \rm kpc$, we find the magnitude of the reflex velocity to be $v_{\rm travel} = 38.6^{+8.3}_{-7.8}\,\rm km \, s^{-1}$.  
Simultaneously, we determine the enclosed MW mass, $M_{\rm MW}(< 50 \, \rm kpc) = 3.36 \pm 0.15 \times 10^{11}\, \rm M_{\odot}$ and the enclosed LMC mass, $M_{\rm LMC}(< 50 \, \rm kpc) = 8.76^{+1.94}_{-1.77} \times 10^{10}\, \rm M_{\odot}$.
Our results suggest that the total LMC mass must be at least $\sim20\%$ that of the MW.
The velocity anisotropy prior to the LMC's infall is constrained to be $\beta_0 = 0.68 \pm 0.02$.
Finally, we demonstrate that neglecting the LMC in models biases the estimated MW mass to prefer more massive values.
\end{abstract}




\section{Introduction}

The mass of the Milky Way (MW) is one of its most fundamental properties, with its value crucial to constructing dynamical models for our Galaxy. 
The MW mass has been estimated through a variety of techniques and visible tracers such as:~stellar streams \citep[e.g.,][]{Gibbons2014, Kupper2015, Malhan2019, Erkal2019, Vasiliev2021, Dillamore2025}, satellite galaxies and globular clusters \citep[e.g.,][]{Zaritsky1989, Vasiliev2019, CorreaMagnus2022}, and MW halo stars \citep[e.g.,][]{Deason2012, Williams2015, Williams2017, McMillan2017, Bird2022, Shen2022, Medina2025}, as well as via dynamics-free approaches \citep[e.g.,][]{Zaritsky2017}.
Most studies converge around constrained values for the virial mass of the MW being $M_{\mathrm{MW}}\sim \,1\times 10^{12}\,\mathrm{M}_{\odot}$ with an uncertainty of approximately $20-30 \%$ \citep[see,][for compilations]{Wang2020, Bobylev2023, Hunt2025}.
Many of these previous estimates have relied on the MW being well described by equilibrium models \citep[e.g., Jeans models,][]{Jeans1915}, i.e., not undergoing a significant merger episode. 

Yet, the MW is currently experiencing a merger with the Large Magellanic Cloud \citep[LMC, see][for a comprehensive review of the effect of the LMC on the MW]{Vasiliev2023} satellite galaxy. 
The LMC is thought to be on its first pericentric passage and to have a dark matter mass $M_{\mathrm{LMC}}\sim 10^{11}\,\mathrm{M}_{\odot}$ \citep[e.g.,][]{Besla2007, Besla2010, Boylan-Kolchin2011, Penarrubia2016, Erkal2019, Koposov2023, Kravtsov2024, Benisty2024, Watkins2024, Brooks2025}.
As the LMC has just completed its most recent pericentric passage at a relative velocity $ > 300 \,\mathrm{km}\,\mathrm{s}^{-1}$ and has a significant mass ratio relative to the MW ($\sim 1:4-10$), the interaction of these two galaxies has induced a state of dynamical disequilibrium throughout the entire MW \citep{Deason2024, Hunt2025}. 
This is also expected and seen within MW--LMC analogues identified within cosmological simulations \citep[e.g.,][]{Darragh-Ford2025, Arora2025, Mansfield2026}.
The infall of the LMC into the MW has generated a density wake in the MW's dark matter halo \citep{Weinberg1998, Garavito-Camargo2019, 2021ApJ...919..109G, Tamfal2021, Rozier2022, Foote2023},
as well as leaving observable signatures in the density and kinematics of stars in the stellar halo \citep[e.g.,][]{Belokurov2019, Conroy2021, Cavieres2025, Chandra2025a, Yaaqib2024, Bystrom2025, Fushimi24, Amarante2024, Sheng2025}.
On large scales, the inner and outer parts of the MW halo have experienced different strengths of acceleration towards the LMC \citep[e.g.,][]{Gomez2015}.
To a Galactocentric observer, the Galactic northern sky appears to be red-shifted and the Galactic southern sky blue-shifted because the halo moves preferentially ‘up’, towards the Galactic north. 
This ‘reflex' displacement manifests as a dipole signal in density \citep{ Garavito-Camargo2021, Conroy2021, Amarante2024} that is higher in the Galactic North, and similarly in stellar radial velocities \citep{2019MNRAS.487.2685E, 2020MNRAS.494L..11P, Petersen2021, Erkal2021, Yaaqib2024, Chandra2025a, Bystrom2025, Brooks2025}.
The magnitude of the reflex velocity dipole is called the \textit{travel velocity} and its on-sky orientation is called the \textit{apex} direction of the reflex motion.
If models of the MW ignore the reflex motion induced by the LMC, then the inference for the MW mass can be overestimated by up to $50\%$ depending on the observed regions of the sky \citep{Erkal2020} .

Simulation Based Inference \citep[SBI, see][for a conceptual overview]{Cranmer2020} is a statistical framework for inference in complex modelling scenarios where traditional analytic methods are impractical. 
Rather than defining a likelihood function, SBI uses many forward simulations to learn the statistical relationship between model parameters and observational data.
Within the astrophysics community, SBI has been increasingly applied to a variety of astrophysics problems \citep[e.g.,][]{Weyant2013, Alsing2019, Jeffrey2021, Lemos2021, Hermans2021, vonWietersheim-Kramsta2024, Lovell2024, Ho2024, Widmark2025, Sante2025, XiangyuanMa2025, Jeffrey2025, Saoulis2025, Nguyen2025, Brooks2025, Brooks2026}.
One of the main probes of the dynamical MW--LMC interaction is the direction and magnitude of the reflex motion. 
Many previous studies have used a likelihood-based approach to constrain this interaction \citep[e.g.,][]{Petersen2021, Yaaqib2024, Bystrom2025, Chandra2025a}, which may lead to biases being introduced due to, e.g., survey selection criteria. 
As a result, SBI offers a novel and flexible method to determine precise measurements of this induced motion, while simultaneously constraining other properties of the MW and LMC \citep{Brooks2025, Brooks2026}.

In \citet{Brooks2026}, we presented and validated a novel SBI architecture for exploring large MW--LMC model parameter spaces to enable rapid and reliable parameter inference of model parameters using the dynamics of outer halo stars. We showed that our SBI framework trained on many rigid MW--LMC simulations retains enough of the relevant physics that more complex simulations capture \citep[e.g., $N$-body simulations,][]{Garavito-Camargo2019} to avoid model misspecification and vastly improve the computational efficiency of the inference. 
Subsequently in \citet{Brooks2025}, we demonstrated the first application of our SBI architecture to data from the Dark Energy Spectroscopic Instrument \citep[DESI,][]{DESICollaboration2025} survey and the H3+SEGUE+MagE \citep{Chandra2025a} combined outer halo survey to constrain the LMC mass, the reflex motion and the strength of dynamical friction. 
One limitation of these studies is that only mean velocities of outer halo stars were used for inference, limiting the ability to constrain the MW mass, which is sensitive to their velocity dispersions \citep[see the spherical Jeans equations,][]{Jeans1915, Binney2008}.
Expanding this SBI framework to incorporate the velocity dispersions of outer halo stars will enable simultaneous constraints to be placed on the MW and LMC masses as well as their reflex interaction.

In this work, we present an enhanced SBI architecture to constrain the MW--LMC interaction. We use both the mean \textit{and} dispersions of outer halo star velocities to place constraints on the MW and LMC masses, the velocity anisotropy parameter, and the reflex motion. 
Additionally, we improve on the previous distance-independent reflex motion parametrisation by implementing a linear continuity model to capture any radial variation of the reflex motion.
Using all-sky kinematics data from the H3+SEGUE+MagE \citep{Chandra2025a} combined with these model enhancements, our SBI framework is able to simultaneously produce precise constraints on the MW and LMC masses, the reflex motion and velocity anisotropy.
Furthermore, with the release of upcoming datasets from Gaia Data Release 4 (DR4) and the Sloan Digital Sky Survey V \citep[SDSS-V,][]{Chandra2025b, SDSSCollaboration2025}, our flexible SBI framework will be uniquely placed to properly account for variation across surveys when performing inference.

The layout of this paper is as follows. 
In Sec.~\ref{sec:data} we describe the data used throughout this work and define the summary statistics used for the inference.
In Sec.~\ref{sec:simulations}, we give a description of the MW--LMC simulations used to train the SBI architecture used for parameter inference.
In Sec.~\ref{sec:sbi}, we describe the SBI architecture, Bayesian statistics and the machine learning models used for inference. 
In Sec.~\ref{sec:results}, we give the constraints on the MW and LMC enclosed and total masses, the reflex motion and the velocity anisotropy model parameters.
In Sec.~\ref{sec:discussion}, we discuss our results in context of previous literature, assess model limitations and investigate the relative importance of including the LMC in models.
Finally, we conclude and provide an outlook for upcoming surveys in Sec.~\ref{sec:conclusions-outlook}.

\section{Data}\label{sec:data}

\subsection{H3, SEGUE \&  MagE Outer Halo Surveys}\label{sec:data-h3-segue-mage}

\subsubsection{H3}\label{sec:data-h3}

The H3 Spectroscopic Survey \citep{Conroy2019} has conducted a spectroscopic survey of halo stars
with the Hectochelle instrument \citep{Szentgyorgyi2011} on the $6.5\rm m$ MMT telescope at the Whipple Observatory in Arizona. 
We use the sample of H3 stars observed up to January 2024 that have reliable stellar parameters from \texttt{MINESweeper} \citep{Cargile2020} and are not associated with known MW substructures. 

\subsubsection{SEGUE}\label{sec:data-segue}

The Sloan Extension for Galactic Understanding and Exploration \citep[SEGUE,][]{Yanny2009} survey observed $\sim250,000$ stars with the low-resolution BOSS spectrograph as a part of the Sloan
Digital Sky Survey \citep[SDSS,][]{York2000}. 
These spectra have been fitted using the \texttt{MINESweeper} routine to provide reliable stellar parameters \citep{Chandra2025b, cargile_2025_16105186}.

\subsubsection{MagE}\label{sec:data-mage}

Over the past two years, a tailored spectroscopic survey of luminous RGB stars in the outer halo has been conducted with The Magellan Echellette Spectrograph \citep[MagE,][]{Marshall2008} on the $6.5\rm m$
Magellan Baade Telescope at Las Campanas Observatory.  
The selection procedure for the target sample of RGB stars is described in \citet{Chandra2023a}, and the details of the spectroscopic survey are further described in \citet{Chandra2023b} and \citet{Chandra2025a}. 
Stellar parameters are estimated with the \texttt{MINESweeper} code \citep{Cargile2020} including parallax measurements from Gaia \citep{GaiaCollaboration2023}.
As of May 2024, a total of $400$ stars have been observed, of which $\sim 300$ are spectroscopically confirmed to be at a heliocentric distance beyond $50 \, \rm kpc$, and $\sim 100$ are beyond $100 \, \rm kpc$ making it the largest dataset of outer halo stars beyond $50 \, \rm kpc$. For a much more detailed account of the \textit{MagE} survey, we direct the reader to \citet{Chandra2023a, Chandra2023b, Chandra2025a}. 

\subsubsection{Combined H3+SEGUE+MagE sample}\label{sec:data-h3-segue-mage-sample}

The above surveys combine to produce a pure sample of stars with homogeneous stellar parameters and their uncertainties derived with the \texttt{MINESweeper} pipeline, e.g., distance estimates with an uncertainty floor of $10\%$. 
Subsequently, we adopt the same selection procedure to exactly reproduce the high-fidelity subset in sec.~2.3, \citet{Chandra2025a}, e.g., fractional distance uncertainties $< 25 \%$.
This selection procedure ensures that known substructures and unphysical quantities are removed from the sample.
In particular, the Sagittarius stream \citep[e.g.,][]{Majewski2003, Vasiliev2021}  is excised using the angular momentum cuts proposed by \citet{Johnson2020}.
We note that specific choices for removing Sagittarius stream stars may alter results, e.g. see appendix A in \citet{Chandra2025a} as well as sec.~2.6 and appendix D in \citet[][]{Bystrom2025}.
The all-sky sample of H3+SEGUE+MagE stars used in this work contains $1,296$ field stars between $r_{\rm gal} \in [30 - 160] \, \rm kpc$. 
The same sample of stars was used in \citet{Brooks2025} when considering the full depth of the H3+SEGUE+MagE combined outer halo survey. Although that study also considered other data selections to match DESI survey selection criteria.
Note that some figures in this study will use the shorthand ‘H3+' to refer to this combined sample.

\begin{figure*}
    \centering
    \includegraphics[width=0.4\linewidth]{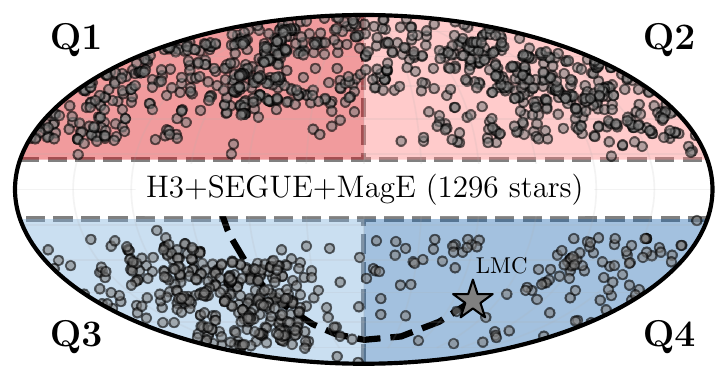}
    \label{figX}
\end{figure*}

\begin{figure}
    \centering
    \includegraphics[width=\linewidth]{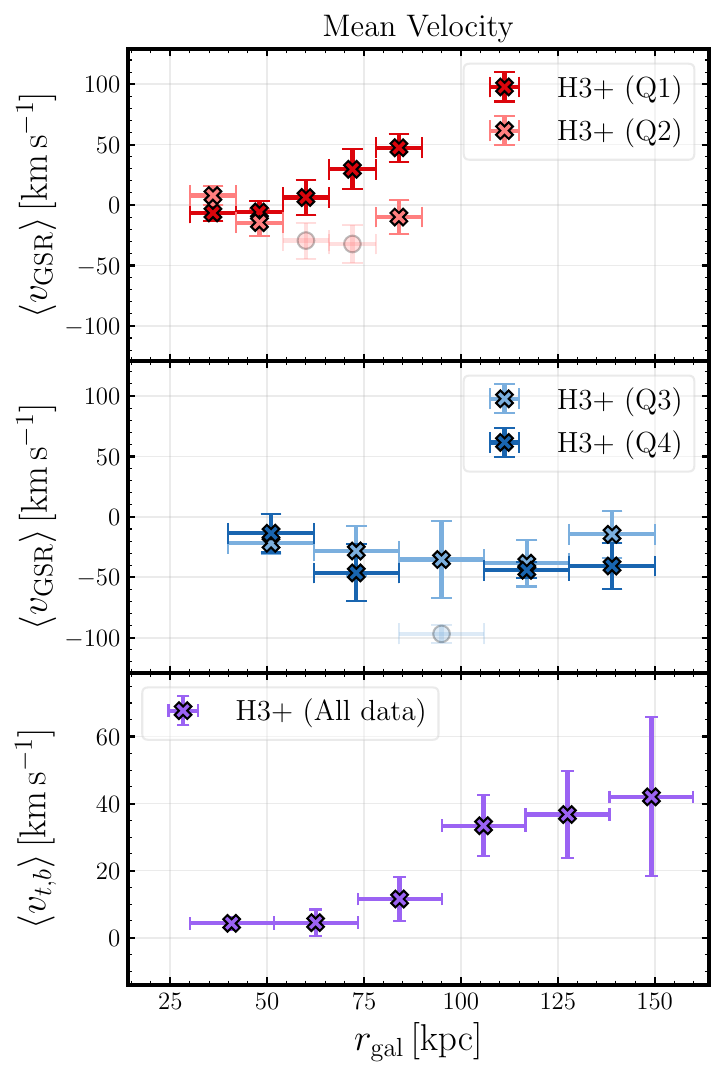}
    \caption{\textbf{Mean velocity distributions} in the radial and tangential components, $\langle{v}_{\rm GSR} \rangle$ and $\langle v_{\rm t, b} \rangle$, as a function of Galactocentric distance for the H3+SEGUE+MagE data divided into on-sky quadrant footprints. Q1/2 are in the Galactic North, Q3/4 are in the Galactic South.
    Vertical error bars represent the $1\sigma$ uncertainties determined via bootstrap resampling. Horizontal error bars display the bin width.
    Data points with increased transparency, and distinct markers, are poorly represented by simulated data counterparts.}
    \label{fig1}
\end{figure}

\begin{figure}
    \centering
    \includegraphics[width=\linewidth]{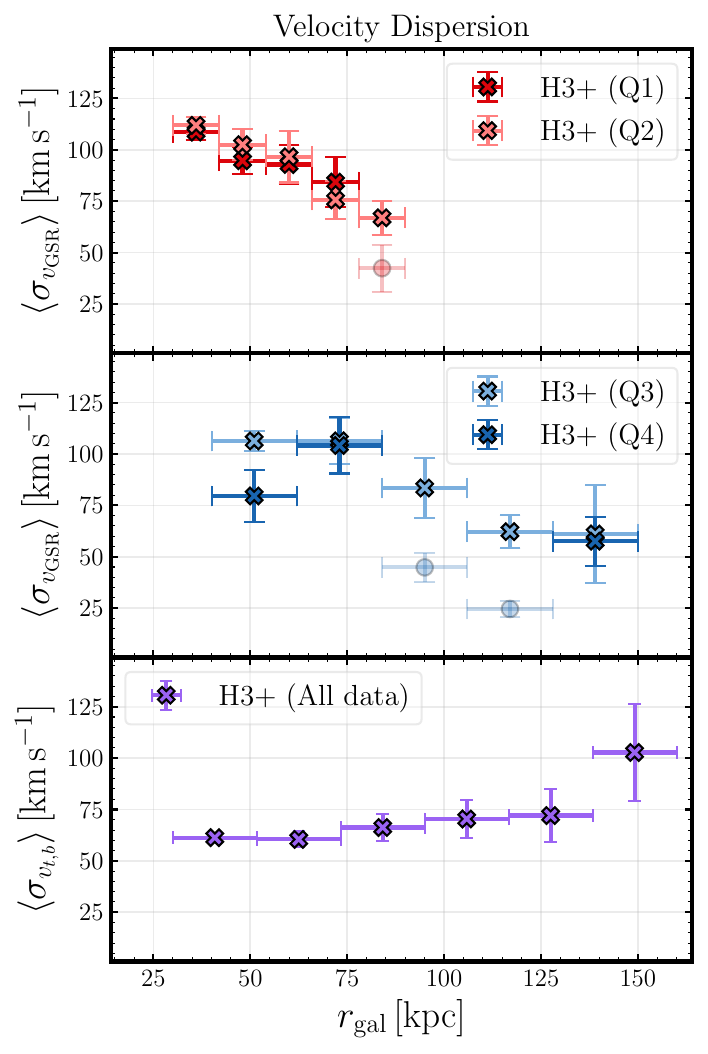}
    \caption{\textbf{Velocity dispersion distributions} in the radial and tangential components, $\langle \sigma_{{v}_{\rm GSR}} \rangle$ and $\langle \sigma_{v_{ t, b}} \rangle$, as a function of Galactocentric distance for the H3+SEGUE+MagE data divided into on-sky quadrant footprints. Q1/2 are in the Galactic North, Q3/4 are in the Galactic South.
    Vertical error bars represent the $1\sigma$ uncertainties determined via bootstrap resampling. Horizontal error bars display the bin width.
    Data points with increased transparency, and distinct markers, are poorly represented by simulated data counterparts.}
    \label{fig2}
\end{figure}

\subsection{Observational summary statistics}\label{sec:data-measurements}

SBI requires a set of summary statistics to be used for the inference process. 
For this application of an SBI framework to the dynamics of the outer MW stars, we adopt summary statistics based on their velocity field distributions. 
Specifically, we focus on the mean and dispersions of the radial and tangential velocities as a function of distance. 
This choice of velocity summary statistics is insensitive to differential selection between the individual H3, SEGUE and MagE surveys.

\subsubsection{Radial velocities}\label{sec:data-measurements-vr}

We calculate the stellar radial velocities in the \textit{Galactic Standard of Rest} (GSR) frame, $v_{\rm GSR}$, which accounts for the solar motion with respect to the Galactic centre. 
For a non-rotating and/or axisymmetric galaxy in equilibrium, the mean radial velocity, $\langle{v}_{\rm GSR} \rangle$, is expected to be zero throughout the entire MW. 
However, our Galaxy is in disequilibrium due to the merger with the LMC and hence $\langle{v}_{\rm GSR} \rangle \neq 0$ ubiquitously across the MW.

We explore the full depth and on-sky coverage of the H3+SEGUE+MagE dataset by dividing the sky into quadrants and measuring a set of mean radial and tangential velocity summary statistics. 
Throughout this work, we define the quadrants as follows: Quadrant 1 (Q1) as $l \in [+180^{\circ}, 0^{\circ}]$, $b \in [0^{\circ}, +90^{\circ}]$, Quadrant 2 (Q2) as $l \in [0^{\circ}, -180^{\circ}]$, $b \in [0^{\circ}, +90^{\circ}]$, Quadrant 3 (Q3) as $l \in [+180^{\circ}, 0^{\circ}]$, $b \in [-90^{\circ}, 0^{\circ}]$ and Quadrant 4 (Q4) as $l \in [0^{\circ}, -180^{\circ}]$, $b \in [-90^{\circ}, 0^{\circ}]$.
A visual representation can be seen in the Mollweide figure above Fig.~\ref{fig1} and Fig.~\ref{fig2}. 
For the all-sky H3+SEGUE+MagE dataset, we calculate the $3 \sigma$-clipped mean, $\langle{v}_{\rm GSR} \rangle$, and dispersions, $\langle \sigma_{v_{\rm GSR}} \rangle$, values for the Galactocentric distance range $r_{\rm gal} \in [30 - 160] \, \rm kpc$ in 5 equally spaced distance bins in the northern quadrants (Q1 and Q2) and the southern quadrants (Q3 and Q4); see the upper and middle panels of Fig.~\ref{fig1} \& Fig.~\ref{fig2}. 
The distance bins are larger than any associated distance uncertainties to minimise correlations between adjacent bins.
Note, that neither the data nor the simulations account for uncertainty in the solar motion with respect to the Galactic centre. 
In these figures, data points with increased transparency, and distinct markers, are those which are poorly represented by simulated data; see Sec.~\ref{sec:results} for more detail.

\subsubsection{Tangential velocities}\label{sec:data-measurements-vt}

The tangential velocity of a star, particularly in the Galactic latitude direction, $v_{t,b}$, traces the LMC’s perturbation in the MW \citep{Erkal2021, Sheng2024, Chandra2025a}. 
This component of the tangential velocity captures the apparent ‘upward’ reflex motion of outer halo stars as the MW's centre of mass is dragged ‘downward’.
We calculate the $3 \sigma$-clipped mean, $\langle{v}_{t,b} \rangle$, and dispersions, $\langle \sigma_{v_{t,b}} \rangle$, values for the Galactocentric distance range $r_{\rm gal} \in [30 - 160] \, \rm kpc$ in 6 equally spaced distance bins; see the lower panels of Fig.~\ref{fig1} \& Fig.~\ref{fig2}. 
These tangential velocities have been corrected for the solar motion and uncertainties determined via bootstrap resampling. 
In this study, we do not consider the tangential velocities in the Galactic longitudinal direction, $v_{t,l}$, though future work can utilise them to constrain the LMC's present-day phase space coordinates and its past orbit \citep{Sheng2025}. 
 
\section{Simulations}\label{sec:simulations}

\subsection{The Milky Way -- LMC potentials}\label{sec:simulations-MWLMCpot}

\subsubsection{The Milky Way}\label{sec:simulations-MW}

To model the MW dark matter halo we use a Navarro-Frenk-White \citep[NFW,][]{Navarro1996, Navarro1997} dark matter halo density profile described by $M_{200}$ and $c_{200}$; see Table.~\ref{table:priors} for the mass prior range considered. 
These quantities are defined by a sphere enclosing an overdensity that is 200 times the critical density of the Universe, $\rho_{\mathrm{crit}} = 3H_{0}^{2}/8\pi G$, as denoted by the ‘$200$' subscript, where the Hubble parameter, $H_0$, is taken to be $67.6\, \rm km\,\rm s^{-1} \, \rm Mpc^{-1}$ using the default cosmology in \texttt{astropy} \citep{astropy:2013}.
We constrain the normalisation of the halo mass profile such that the circular velocity at the solar position is approximately $235\,\mathrm{km}\,\mathrm{s}^{-1}$ \citep[e.g., matching constraints from][within the associated uncertainty]{McMillan2017}.\footnote{Also see \citet{Horta2026} for a recent empirical measurement using element abundance gradients.}
The MW stellar components are modelled using a spherical bulge with a total mass $1.2 \times 10^{10} \, \mathrm{M}_{\odot}$, and an exponential stellar disc with a total mass $5 \times 10^{10} \, \mathrm{M}_{\odot}$. 
The stellar distributions remain fixed taking the values suggested by \cite{McMillan2017}. 

\subsubsection{The LMC}\label{sec:simulations-lmc}

We model the LMC as a Hernquist dark matter halo \citep{Hernquist1990}. 
We normalise the profile such that the derived rotation curve peaks at $(91.7 \pm \, 18.8) \,\mathrm{km}\,\mathrm{s}^{-1}$ at a distance of $8.7\,\mathrm{kpc}$ from its centre. This corresponds to an enclosed dynamical mass of $M_{\rm LMC}(r<8.7\,\mathrm{kpc}) = (1.7 \pm 0.7) \times 10^{10}\,\mathrm{M_{\odot}}$ \citep{vanderMarel2014}. 
See Table.~\ref{table:priors} for the chosen total LMC mass prior distribution.

\subsubsection{The MW--LMC interaction}\label{sec:simulations-mwlmcinteraction}

The trajectories of the MW and LMC under their mutual gravitational attraction are numerically integrated \citep[see equ.~3-6 in][]{Brooks2026}. 
We account for acceleration due to Chandrasekhar dynamical friction, $\mathbf{a}_{\mathrm{DF}}$, on the trajectory of the LMC \citep{Chandrasekhar1943, Binney2008, Jethwa2016} as:

\begin{equation}\label{equ1}
\mathbf{a_{\rm DF}} = - \frac{4 \pi G^2 M_{\rm LMC} \rho_{\rm MW} \ln \Lambda}{v_{\rm LMC}^3} 
\left[ \operatorname{erf}(X) - \frac{2X}{\sqrt{\pi}} e^{-X^2} \right]  \lambda_{\mathrm{DF}} \, \mathbf{v_{\rm LMC}},
\end{equation}
\noindent
where $X = v_{\rm LMC} / \sqrt{2}\sigma_{\rm MW}$ and $\sigma_{\rm MW}$ and $\rho_{\rm MW}$ are the velocity dispersion and total density field of the MW. 
Following \citet{Vasiliev2021}, we take a fixed value of $\sigma_{\rm MW} = 120 \,\rm km \, \rm s^{-1}$ for the velocity dispersion as the dynamical friction is insensitive to the precise value. 
For the Coulomb logarithm we adopt \(\ln \Lambda = \ln \left( 100\, \rm kpc / \epsilon \right)\). The softening length, $\epsilon$, depends on the satellite’s density profile \citep{White1976}. 
We adopt $\epsilon = 1.6 \, a_{\rm LMC}$, where $a_{\rm LMC}$ is the LMC scale radius, as this has been used previously when modelling the LMC as a Plummer sphere \citep[e.g.,][]{Hashimoto2003, Besla2007, vanderMarel2012, Sohn2013, Kallivayalil2013, Patel2020}. 
The numerator in the Coulomb logarithm expression is an arbitrarily chosen value that loosely describes the average separation of the MW and LMC. 
This value could be updated through the integration of the LMC orbit, though it would have a small effect. 
We use a dimensionless parameter, $\lambda_{\mathrm{DF}}$, to modulate the strength of the dynamical friction that the LMC experiences. 
In principle, this will take into account changes to the fixed Coulomb logarithm value per simulation. 

\subsection{The Milky Way Stellar Halo}\label{sec:simulations-stellarhaloes}

To generate a mock MW stellar halo for each simulation, we draw phase-space samples from radially-biased distribution functions as implemented in \textsc{agama} \citep{2019MNRAS.482.1525V}.
This requires instances of a tracer density profile, a potential, and a prescription for the radial velocity anisotropy. 
We use a spheroidal tracer density profile,

\begin{equation}
\rho  \propto  \left( \frac{\tilde{r}}{r_{\rm dens}} \right)^{-1}
\left[ 1 + \left( \frac{\tilde{r}}{r_{\rm dens}} \right) \right]^{1 - \gamma},
\end{equation}
\noindent
allowing the logarithmic slope of the outer halo to vary between $\gamma \in [3, 5]$\footnote{Although $\beta$ is often used to denote the outer density slope, we adopt $\gamma$ here to avoid confusion with the velocity anisotropy parameter $\beta$.} to account for the range of previous literature values \citep[e.g.,][]{Han2022, Amarante2024, Cavieres2025, Li2025}. 
We note that the inner slope is fixed, $\rho \propto r^{-1}$, making it formally inconsistent with very radially biased distributions via the \textit{cusp slope --  anisotropy theorem} \citep{An2006}. 
In this study, we use stars well beyond the scale radius to perform parameter inference, making this inconsistency negligible here.
For the potential, we use the combined rigid MW profile, consisting of an NFW dark matter halo, a stellar disc and bulge \citep{Navarro1996, Navarro1997} with relevant parameters adopted from each unique MW--LMC simulation; see, Sec.~\ref{sec:simulations-MW}.  
We implement the radial bias of stellar velocities \citep{Osipkov1979, Merritt1985, Binney2008} using a constant velocity anisotropy profile such that the anisotropy remains constant and equal to $\beta_0$ over all radial scales.

We sample $20,000$ phase-space coordinates per stellar halo to minimise Poisson noise, ensuring observational uncertainty dominates. This precision allows us to convolve simulated data with survey-specific observational uncertainties to produce observational-like quantities used for the inference.
In this study, we apply the survey-specific uncertainties to the binned radial and tangential velocity measurements from the H3+SEGUE+MagE dataset; see Sec.~\ref{sec:data-measurements}. 
This approach allows us to forward model all the stellar haloes to match a specific survey of interest, in this case the combined H3+SEGUE+MagE outer halo survey, and perform the subsequent inference.
For a given MW--LMC potential with reflex motion, we integrate all particles in the stellar halo as test particles to present-day over the last $2.2\,\mathrm{Gyr}$ \citep[motivated by the epoch when the LMC crosses the MW virial radius in][]{Garavito-Camargo2019}. 

\subsubsection{The reflex motion}

From the final distribution of stellar halo particles, we measure the reflex motion of the MW in response to the infalling LMC \citep[as used in][]{Petersen2021, Chandra2025a, Yaaqib2024, Yaaqib2025, Brooks2025, Brooks2026}.
This method fits an on-sky velocity model which contains nine free parameters. 
We model the dipole reflex motion using Galactocentric Cartesian velocities $\{v_x, v_y, v_z \}$, which can be transformed to spherical coordinates $\{v_r, v_{\phi}, v_{\theta} \}$ to combine with the mean motions. 
We choose to sample in Cartesian coordinates rather than spherical coordinates to avoid inefficient convergence during maximum likelihood estimation if the preferred apex direction is close to the Galactic poles.
We account for non-zero mean motion in the halo's Galactocentric velocity via the mean motion parameters $\vec{v}_{\mathrm{mean}} = \left( \langle v_r \rangle, \langle v_{\phi} \rangle, \langle v_{\theta} \rangle \right)$. This allows for any departures in the bulk halo motion from the travel velocity.
Finally, we account for the intrinsic velocity dispersion in each component using the set of parameters $\{ \sigma_{v_{\rm r}},\sigma_{v_l}, \sigma_{v_b} \}$.
We note that within our SBI framework, these parameters act as nuisance parameters, since measurement uncertainties are incorporated into the velocity summary statistics used for inference; see Sec.~\ref{sec:data-measurements}. 
This parametrisation of the intrinsic velocity dispersion is only used to obtain a maximum likelihood estimate of the ‘true' reflex motion parameters for each simulation.
The reflex motion model is represented by the sum of the travel velocity and mean velocity parameters:
\begin{equation}
    \langle \vec{v}\rangle = \vec{v}_{\rm travel} + \vec{v}_{\mathrm{mean}}, 
\end{equation}
\noindent
where $\langle \vec{v} \rangle$ is the mean Galactocentric halo velocity vector. 
To find the maximum likelihood estimates for these parameters given each set of mock stellar halo data, we minimise a Gaussian log-likelihood for the 1-dimensional line of sight velocities and 2-dimensional proper motions using \texttt{scipy.optimize} \citep[see equations 6 \& 8 in][]{Petersen2021}. 
We return the maximum likelihood estimates for all of the reflex motion model parameters. 
For this study, we will comment on the magnitude, $v_{\rm travel}$ or Galactocentric components of the reflex travel velocity, $\{v_x, v_y, v_z \}$ and the apex direction, $(l_{\rm apex}, b_{\rm apex})$.

We improve upon our distance-independent model in \citet{Brooks2026, Brooks2025} by employing a reflex motion model that is allowed to vary linearly as a function of Galactocentric distance \citep[see a similar approach in][]{Chandra2025a}. 
To achieve this, the reflex motion model now takes unique parameter values at $r_{\rm gal} = 40 \, \rm kpc$ and $r_{\rm gal} = 120 \, \rm kpc$, with a linear fit implemented between those two anchoring values. 
Overall, the number of reflex motion model parameters is doubled. 
This simple linear expansion allows us to find simulation prior distributions for reflex motion model parameters at any distance between $r_{\rm gal} \in [40, 120] \, \rm kpc$, and allows for tomographic inference of the reflex motion; see Sec.~\ref{sec:results-setup}.

\subsection{Simulation Priors}\label{sec:simulations-ics}

We update some of the parameter prior probability distributions from \citet{Brooks2025, Brooks2026}. We show all of our priors in Table.~\ref{table:priors}.
The first two parameters are the MW mass and LMC total mass.
Next is $\lambda_\mathrm{DF}$, the scalar multiple that modulates the strength of dynamical friction relative to classic Chandrasekhar values; see Equ.~\ref{equ1}.
The next set of parameters describe the LMC present-day position and velocity with their distributions inspired by the values in sec.~3.1 and table~2 of \citet{Vasiliev2023}. 
The final three parameters are the anisotropy parameter, $\beta_0$, the scale length, $r_{\rm dens}$, and outer halo slope, $\gamma$, for the tracer density profile that initialise the mock MW stellar haloes.
In total, we run $32,000$ MW--LMC simulations each with unique parameter values and $20,000$ particles to represent the MW stellar halo.

\begin{table}
\centering
\caption{Simulation/model parameter prior distributions.}
\label{table:priors}
\begin{tabularx}{\linewidth}{lX}
\hline
\hline
Model Parameter & Prior probability distribution \\
\hline
\input{tables/priors.tex}
\end{tabularx}
\end{table}

\section{Simulation Based Inference}\label{sec:sbi}

In the Bayesian approach, a problem is often posed as calculating the probability of the model
parameters, $\theta$ (e.g., $M_{200, \rm MW}$), given some observed data, $D_{\text{obs}}$ (e.g., $\langle v_{\rm GSR} \rangle$), and a theoretical model, $I$ (e.g.,  rigid MW--LMC simulations). 
In other words, we want to find the posterior probability distribution, $\mathcal{P} = p(\theta | D_{\text{obs}}, I)$. This is possible using Bayes' Theorem:

\begin{equation}\label{equ4}
     p(\theta | D_{\text{obs}}, I) = \frac{p(D_{\text{obs}} | \theta, I) p(\theta | I)}{p(D_{\text{obs}} | I)}
\iff \mathcal{P} = \frac{\mathcal{L} \times \Pi}{\mathcal{Z}} 
\end{equation}
\noindent
where $\mathcal{L} = p(D_{\text{obs}} | \theta, I)$ is the likelihood, $\Pi =   p(\theta | I)$ is the prior, and $\mathcal{Z} = p(D_{\text{obs}} | I)$ is the
Bayesian evidence. The Bayesian evidence acts as a normalisation in parameter estimation and can be ignored for our application. Given a choice of prior distribution for parameters and a likelihood function, we can find the posterior distribution. SBI allows an estimate for the likelihood, and hence the posterior, distributions to be found without the need for assumptions about the analytic form for the likelihood.

The roots of SBI lie within the Approximate Bayesian Computation framework \citep[ABC, e.g.,][]{Rubin1984, Pritchard1999, Fearnhead2010}. An ABC framework selects forward simulations that are the most similar to the observed data based on some distance measure involving summary statistics of the simulation.
Another way to compute the posterior is via Density Estimation Likelihood Free Inference (DELFI). In this approach, forward simulations are used to learn a conditional density distribution of the data $D_{\text{obs}}$, given the simulation parameters $\theta$, using a density estimation algorithm, e.g., normalising flows, that utilise a series of bijective transformations to convert a simple base distribution e.g., a Gaussian, into the target probability distribution \citep{JimenezRezende2015}. 

We adopt the DELFI approach as implemented in the \texttt{sbi}\footnote{Also available via the \texttt{ltu-ili} Python package which unifies several of the available SBI codes \citep{Ho2024}} Python package \citep{tejero-cantero2020sbi} to directly estimate the posterior distribution; often called \textit{neural posterior estimation}.
The specific choice of neural network architecture used can influence the estimated posterior. 
Firstly, the choice of modelling hyperparameters, e.g., the number of layers and nodes defining the neural network, can affect the quality of the trained SBI model. 
We find that a Masked Autoregressive Flows \cite[MAF,][]{Papamakarios2017, Papamakarios2019} architecture produces well calibrated posteriors using $5$ transformation layers in the neural network, each with a width of $50$ nodes.  
Secondly, we account for the model (epistemic) uncertainties, i.e., those stemming from the uncertainty in the choice of network weights and biases, by model ensemble averaging.

DELFI is advantageous over the simpler ABC approach as it does not rely on a choice of a distance measure and it uses all available forward simulations to build the posterior distribution, making it far more efficient \citep{Alsing2019}. 
Once a neural posterior estimator has been trained on a precomputed simulation dataset, a posterior can be returned for many unique observations without having to retrain the flow; this is known as \textit{amortisation} \citep{Mittal2025}.

The simulations used for SBI in this work are described in Sec.~\ref{sec:simulations}. 
We use a MAF density estimator to directly obtain the posterior distribution that can be evaluated at any observed data point for any data realisation, i.e., \( p(\theta | D_{\text{obs}}, I) \). In general, directly estimating the posterior is preferred over estimating the likelihood when a neural network maps high-dimensional inputs to low-dimensional outputs \citep{Ho2024}.
We ensure the reliability of the estimated posteriors through some diagnostic checks including coverage probabilities and predictive posterior checks in Appendix \ref{sec:posterior-checks}.

\section{Results}\label{sec:results}

In this section we show posterior distributions for MW--LMC model parameters conditioned on different subsets of the outer halo H3+SEGUE+MagE survey.
All figures use $10,000$ samples drawn from their respective posterior density distributions.

\subsection{Inference set-up}\label{sec:results-setup}

We will show the posterior distributions for MW and LMC masses enclosed within $50\,\rm kpc$, the Galactocentric Cartesian components of the reflex motion and the velocity anisotropy parameter. 
Additionally, using these posterior samples, we will show the magnitude and on-sky apex direction of the reflex motion, as well as the radial variation in these reflex motion parameters.

For the input data that is used to condition the posterior estimation on, 
we use the mean and dispersions of the radial and tangential velocities from the outer halo H3+SEGUE+MagE survey; see Fig.~\ref{fig1} and Fig.~\ref{fig2}. 
To investigate the effect of varying levels of velocity data provided we use various subsets of this velocity data.

\begin{description}
    \item[\textbf{Yellow contours,} H3+ $\langle v \rangle$:] The entire mean velocity data is used to condition the posterior estimation, i.e., all data points in Fig.~\ref{fig1}.
    \vspace{0.2cm}
    \item[\textbf{Blue contours,} H3+ $\langle v \rangle, \, \langle \sigma_v \rangle$:] The entire mean and dispersion velocity data is used to condition the posterior estimation, i.e., all data points in Fig.~\ref{fig1} and Fig.~\ref{fig2}. 
    \vspace{0.2cm}
    \item[\textbf{Red contours,} H3+ $\langle v \rangle^{*}, \, \langle \sigma_v \rangle^{*}$:] Only the mean and velocity dispersion data points which are well represented by our simulations are used to condition the posterior estimation. Specifically, we remove all data points that lie $2\sigma$ or more away from the corresponding simulated data mean. These ‘discrepant' data points are indicated by the transparent markers in Fig.~\ref{fig1} and Fig.~\ref{fig2}. Overall, $6/52$ data points are removed.
\end{description}

\noindent
The removal of discrepant data allows us to investigate any bias introduced by including data points which are not well represented by the current simulations. This effect can be two-fold. On one hand, the observational dataset itself could still contain undiagnosed substructure producing outlier values. On the other hand, the rigid simulation set-up may not contain enough of the relevant physics \citep[e.g.,][]{Yaaqib2025} to truly represent these data points; see Sec.~\ref{sec:discussion} for further discussion on the origin of discrepancies between the simulations and data.
This is especially important to consider when using an SBI approach to perform parameter inference as the simulations must be able to properly represent the observed data.
As a reminder, observable uncertainties are accounted for in our simulated stellar halo measurements; see Sec.~\ref{sec:simulations-stellarhaloes}  for details on accounting for survey specific uncertainties.
Finally, we adopt the present-day Galactocentric position and velocity of the LMC to be $\vec{x}_{\rm LMC} = 
\{-0.6, -41.3, -27.1 \} \, \rm kpc$, $\vec{v}_{\rm LMC} = \{-63.9, -213.8, 206.6 \} \, \rm km \, s^{-1}$ \citep[][and references therein]{Vasiliev2021} and the outer slope of the stellar halo to be $\gamma=4.55$  \citep[global stellar halo fit,][]{Amarante2024}.
We note that both the LMC present-day coordinates and the outer halo slope are model parameters explored in their respective prior spaces. 
We choose to condition our inference on these values as they are well motivated and are not the main focus of this work.
Future studies will use this SBI framework to constrain these model parameters. 

\subsection{The mass of the Milky Way and Large Magellanic Cloud}

\begin{figure}
    \centering
    \includegraphics[width=\linewidth]{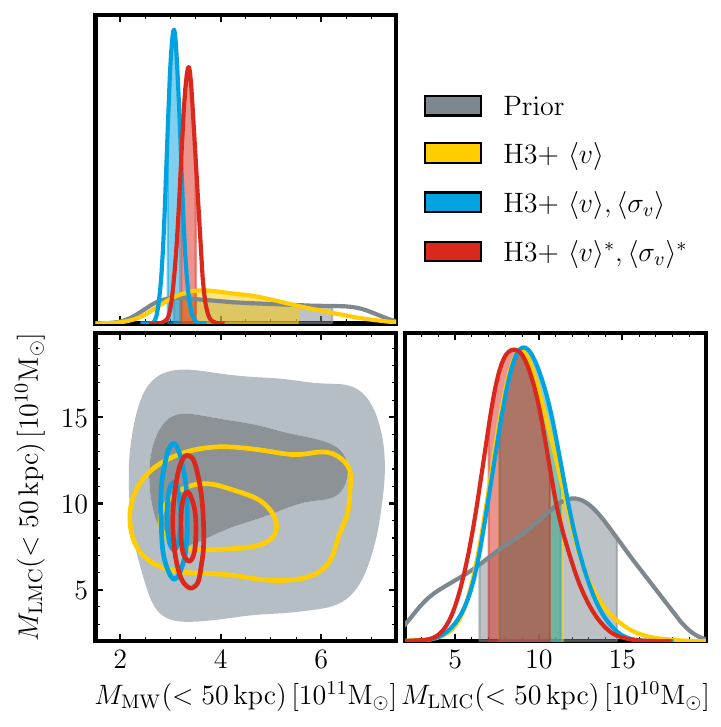}
    \caption{\textbf{The MW and LMC enclosed masses:} The joint, and individual, posterior distributions for the MW and LMC masses enclosed within $50\,\rm kpc$ 
    using data from the H3+SEGUE+MagE survey.
    The yellow and blue contours represent the posteriors conditioned using the means, or means \textit{and} dispersions, of the stellar velocities, respectively.
    The red contours use the mean and dispersion data, but removing discrepant data points.
    The prior distributions are shown as the filled grey contours.
    The contours delineate the $1\sigma$ and $2\sigma$ confidence intervals. 
    For the 1D posterior panels we show the $16^{\rm th}-84^{\rm th}$ percentiles as shaded regions.}
    \label{fig3}
\end{figure}

Previously, in \citet{Brooks2025} we used the H3+SEGUE+MagE dataset and SBI to constrain the enclosed mass of the MW and LMC. In that work, only the means of the radial and tangential velocities were used for the inference. 
In this study, we use the velocity means \textit{and} dispersions to perform the inference. 
By conditioning the posteriors on both of these velocity moments we are able to constrain the MW mass to a much greater precision \citep[e.g.,][]{Jeans1915, Binney2008, CorreaMagnus2022, Sheng2025}.
In Fig.~\ref{fig3} we show the posterior distributions for the MW and LMC masses enclosed within $50\,\rm kpc$ from this study. 
The yellow and blue contours represent the posteriors conditioned using the means, or means \textit{and} dispersions, of the stellar velocities, respectively.
The red contours represent the posterior conditioned on the mean and dispersion data after removing any discrepant data points.
The contours delineate the $1\sigma$ and $2\sigma$ confidence intervals. 
For the 1D posterior panels we show the $16^{\rm th}-84^{\rm th}$ percentiles as shaded regions.
The prior distributions are shown as the filled grey contours. Note that these priors differ from those used in the previous analysis of this dataset \citep{Brooks2025}. We choose here to implement broader and less informative mass priors to reduce the influence of the choice of an informative prior on our results.

Using only the mean velocities, i.e., the yellow contours in Fig.~\ref{fig3}, we constrain the LMC mass enclosed to be $M_{\rm LMC}(<50\,\rm kpc) = 9.31^{+2.17}_{-1.78} \times 10^{10} \, \rm M_{\odot}$, in agreement with our previous study using the same dataset and a similar SBI set-up \citep[][]{Brooks2025}. 
We do not report a total mass ratio here as the MW mass posterior is prior dominated.

Using the full set of mean and velocity dispersion measurements to condition the posterior estimation on, i.e., blue contours in Fig.~\ref{fig3}, we constrain the MW enclosed mass to be $M_{\rm MW}(<50\,\rm kpc) = 3.07 \pm 0.13 \times 10^{11} \, \rm M_{\odot}$ and the LMC enclosed mass as $M_{\rm LMC}(<50\,\rm kpc) = 9.37^{+1.94}_{-1.79
} \times 10^{10} \, \rm M_{\odot}$. The inclusion of this velocity dispersion data helps to constrain the MW mass \citep[][]{Sheng2025}.

Finally, after removing any discrepant observational data,  i.e., red contours in Fig.~\ref{fig3}, we constrain the MW enclosed mass to be $M_{\rm MW}(<50\,\rm kpc) = 3.36 \pm 0.15 \times 10^{11} \, \rm M_{\odot}$ and the LMC enclosed mass as $M_{\rm LMC}(<50\,\rm kpc) = 8.76^{+1.94}_{-1.77} \times 10^{10} \, \rm M_{\odot}$.
Hence, including data which is not well captured by the current rigid MW--LMC simulations can bias the inference in the MW mass by $\sim10\%$. 
As expected, the MW mass inference favours lower values when using all data points because excluded points have lower velocity dispersions, which prefer lower masses \citep{Jeans1915}.
Moreover, we find that the inference of the total masses for the MW, $M_{200, \rm MW}$ and LMC, $M_{\rm LMC}$, gives a mass ratio of $\approx30 \pm 10 \%$ for the LMC's infall mass to the MW's present day mass. 

\subsection{The reflex motion}

\begin{figure*}
    \centering
    \includegraphics[width=\linewidth]{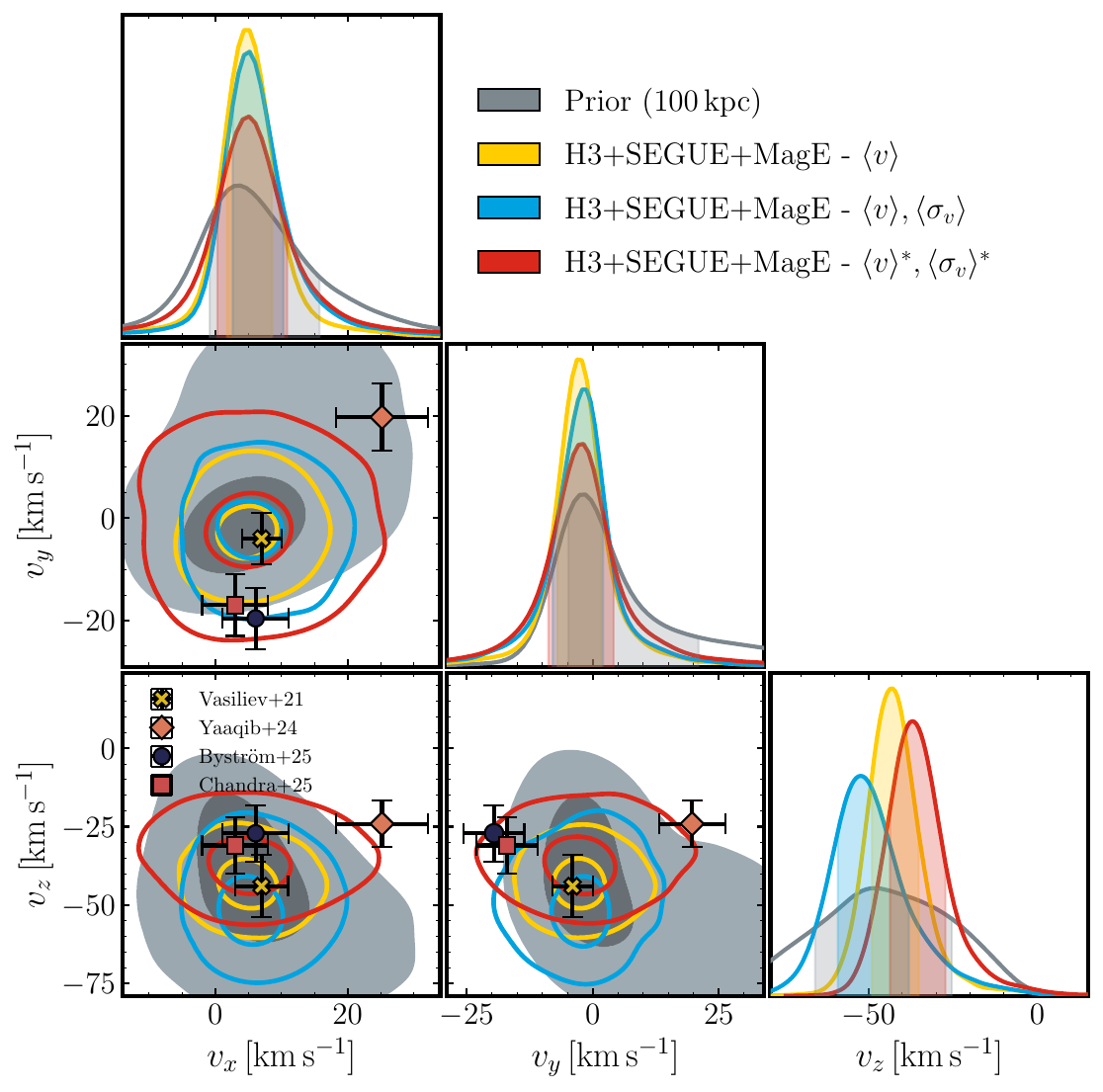}
    \caption{\textbf{The reflex motion velocity:} The joint, and individual, posterior distributions of the Galactocentric Cartesian travel velocity components
    using data from the H3+SEGUE+MagE survey.
    The yellow and blue contours represent the posteriors conditioned using the means, or means \textit{and} dispersions, of the stellar velocities, respectively.
    The red contours use the mean and dispersion data, but removing discrepant data points.
    The contours delineate the $1\sigma$ and $2\sigma$ confidence intervals.
    For the 1D posterior panels we show the $16^{\rm th}-84^{\rm th}$ percentiles as shaded regions.
    The measured mean and $1\sigma$ errors from \citet[][yellow cross]{Vasiliev2021}, \citet[][orange diamond]{Yaaqib2024}, \citet[][blue circle]{Bystrom2025} and \citet[][red square]{Chandra2025a} are shown in each panel for comparison.}
    \label{fig4}
\end{figure*}

\begin{figure*}
    \centering
    \includegraphics[width=0.55\linewidth]{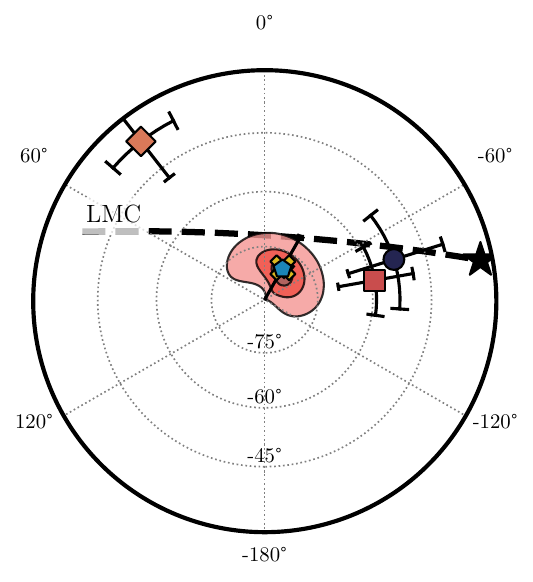}
    \includegraphics[width=0.8\linewidth]{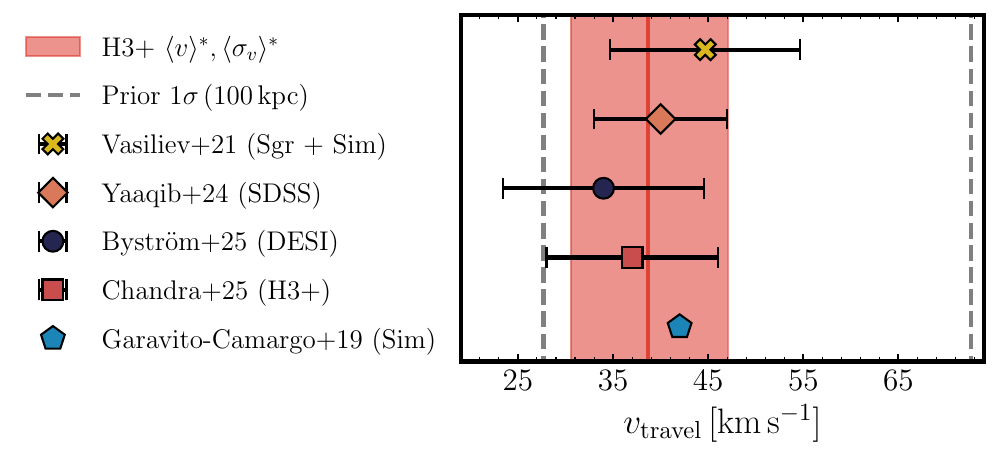}
    \caption{\textbf{Apex direction and magnitude of the reflex motion --} 
    Top panel: A stereographic projection centred on the south Galactic pole showing the apex direction of the travel velocity, $(l_{\rm apex}, b_{\rm apex})$. 
    The red posterior contours are placed at the $40^{\rm th}$, $70^{\rm th}$, and $90^{\rm th}$ quantiles, corresponding to Gaussian-equivalent levels of $0.5 \sigma$, $1 \sigma$, and $1.5 \sigma$. 
    The posterior distribution is conditioned on the mean \textit{and} dispersion velocity data after removing discrepant data points.
    The values from existing literature are shown as markers \citep[][in yellow using Sgr data, orange using SDSS data, red using H3+SEGUE+MagE data, and dark blue using DESI data, respectively]{Vasiliev2021, Yaaqib2024, Chandra2025a, Bystrom2025}. 
    Additionally, we show the apex direction of the fiducial $N$-body simulation in \citet{Garavito-Camargo2019} as the light blue pentagon.
    The past LMC orbit is calculated using the best-fit values from the inference with the present-day location denoted by the black star.
    Lower panel: A comparison of the magnitude of the travel velocity, $v_{\rm travel}$, from the same literature values and our posterior median and $1 \sigma$ percentiles as the red line and shaded region.
    The prior $1\sigma$ confidence interval is shown as the grey dashed lines.}
    \label{fig5}
\end{figure*}

Previously, in \citet{Brooks2025} we reported distance-independent results for the reflex motion parameters using the same H3+SEGUE+MagE dataset.
In this study, we have implemented a linear continuity model, see Sec.~\ref{sec:simulations-stellarhaloes}, to capture the radial dependence in the reflex motion parameters \citep{Yaaqib2024, Chandra2025a}.
As a reminder, the linear continuity model uses the information of all stars between $40\,\rm kpc$ and $120\,\rm kpc$.
We provide tabulated posterior results at $40\,\rm kpc$, $100\,\rm kpc$ and $120\,\rm kpc$ in Table.~\ref{table:reflex-post} for the magnitude, $v_{\rm travel}$, on-sky apex direction $(l_{\rm apex},b_{\rm apex})$ and then Galactocentric Cartesian components, $(v_x, v_y, v_z)$ of the reflex motion.
In Fig.~\ref{fig4} and Fig.~\ref{fig5} we report the reflex motion posteriors evaluated at a reference distance of $100\,\rm kpc$.

In Fig.~\ref{fig4} we show the posterior distribution of the Galactocentric velocity components of the reflex motion.
Again, we show the posterior distributions conditioned on varying subset of data as the coloured contours; see Sec.~\ref{sec:results-setup} for explicit details.
The prior distributions are shown as the filled grey contours, covering a much broader region of this parameter space compared to the distance-independent approach previously used \citep{Brooks2025}. 
We show the measured median values from \citet[][yellow cross]{Vasiliev2021}, \citet[][SDSS/SEGUE data, $r_{\rm gal}>50 \, \rm kpc$ distance bin, orange diamond]{Yaaqib2024}, \citet[][DESI data, blue circle]{Bystrom2025} and \citet[][H3+SEGUE+MagE data, best-fit value at $100\,\rm kpc$, red square]{Chandra2025a} in each panel.
Although not unexpected, the addition of the velocity dispersion data does not provide any extra constraining power. 
In all velocity components, our result agrees with  \citet{Vasiliev2021} within $1\sigma$ and with \citet{Bystrom2025, Chandra2025a} within $2\sigma$. However, our result is very discrepant compared to \citet{Yaaqib2024}.
Interestingly, \citet{Vasiliev2021} is the only other displayed reflex motion result that is constrained by fitting a simulation of the MW--LMC system to data. However, they use a completely different data set of Sagittarius stream stars rather than outer Milky Way halo stars.
In \citet{Yaaqib2024, Bystrom2025, Chandra2025a} the components of the travel velocity are constrained as free parameters, directly fitting the observations rather than an assumed MW--LMC model. 
The agreement between studies that fit simulation models to the reflex motion \citep[i.e., this study and][]{Vasiliev2021} should not be interpreted as a methodological success; rather, this indicates that the simulation models used share similar biases. 
Moreover, the MW is likely much more complex in reality. 
Indeed, this can be seen via the disagreement between studies directly fitting the data \citep[i.e.,][]{Yaaqib2024, Bystrom2025, Chandra2025a}.

In Fig.~\ref{fig5} we show the posterior distribution of the reflex motion described by the magnitude of the velocity dipole vector, $v_{\rm{travel}}$, and its \textit{apex} direction\footnote{See this \href{https://gist.github.com/dc-broo3/f246dc4824b43d6a20b8a122f9d29a92}{tutorial} for a reproducible version of the stereographic projection showing the apex direction in Fig.~\ref{fig5}.} of the reflex motion $(l_{\rm{apex}}$, $b_{\rm{apex}})$ in Galactic coordinates \citep[e.g.,][]{Vasiliev2021, Petersen2021, Yaaqib2024, Bystrom2025, Chandra2025a}. 
The coloured contours and markers represent the same information as in Fig.~\ref{fig4}.
Interestingly, the fiducial $N$-body simulation of the MW--LMC \citep[see blue pentagon,][MWLMC5 model]{Garavito-Camargo2019} is also inconsistent with these previous observations, but agrees within the $1\sigma$ uncertainty of our posteriors.
This suggests that modelling the longitudinal apex direction, $l_{\rm apex}$, remains a challenge even for higher fidelity simulations of the MW--LMC system. 
Nevertheless, we find a good agreement between our results and the existing constraints for the magnitude of the reflex motion. 
The result found using the mean \textit{and} dispersion velocity data after removing discrepant data points produces a travel velocity of  $v_{\rm travel} = 38.6^{+8.3}_{-7.8} \, \rm km \, \rm s^{-1}$ and apex direction of $(l_{\rm apex} [^{\circ}], b_{\rm apex} [^{\circ}]) = (-20.0^{+58.4}_{-51.0}, -75.9^{+20.5}_{-7.0})$ at a reference distance of $100\,\rm kpc$.

\begin{figure}
    \centering
    \includegraphics[width=\linewidth]{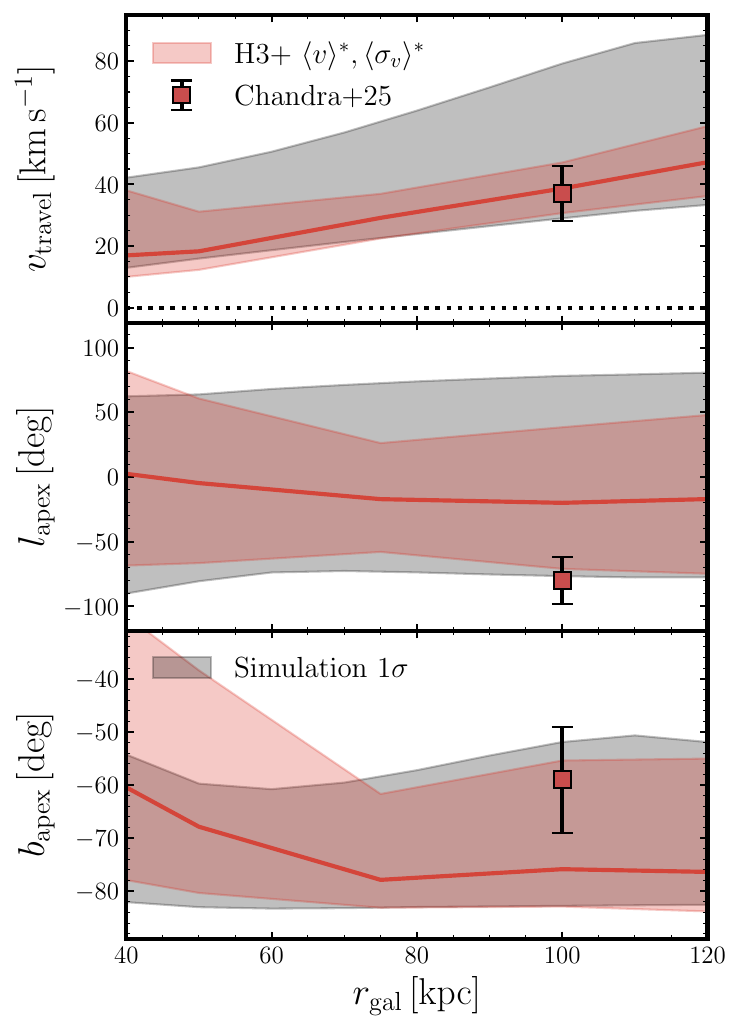}
    \caption{\textbf{Radial variation of the reflex motion:} The radial variation of the magnitude, $v_{\rm travel}$, and the apex direction, $(l_{\rm apex}, b_{\rm apex})$, of the reflex motion posteriors.
    We show the median as the red line and the $1\sigma$ confidence interval as the shaded band.
    The simulation prior $1\sigma$ confidence interval is shown as the grey shaded band.
    The red data point represents the result from \citet{Chandra2025a} evaluated at $100 \, \rm kpc$ using the same H3+SEGUE+MagE dataset but distinct methodology. }
    \label{fig6}
\end{figure}

In Fig.~\ref{fig6}, we explore the radial variation of the reflex motion. 
In the top panel, we show the distance dependence for the travel velocity posterior conditioned on the mean and dispersion velocity data, having removed discrepant data points, as the red median and shaded $1\sigma$ band. 
The simulation prior $1\sigma$ band is shown by the grey shaded region. 
There is a clear increase in the travel velocity from $v_{\rm travel}(r_{\rm gal} = 40  \, \rm kpc) = 17.0^{+21.2}_{-7.0} \, \rm km \, s^{-1}$ to $v_{\rm travel}(r_{\rm gal} = 120  \, \rm kpc) = 47.2^{+11.7}_{-11.0} \, \rm km \, s^{-1}$.
This confirms the measured growth in the travel velocity as a function of distance as seen in \citet{Chandra2025a} using the same dataset and linear continuity model. 
Moreover, \citet{Chandra2025a} demonstrated that using a linear model is sufficient to capture the overall radial dependence of the reflex motion parameters.
We find that the result in this work and \citet{Chandra2025a} evaluated at a reference distance of $100\,\rm kpc$ agree within $1\sigma$ uncertainties.
In the middle and bottom panels, we show the radial dependence of the apex parameters. 
Similarly, we find there is little radial variation in the apex direction, as seen previously in the data \citep[see fig.~7 in][]{Chandra2025a}, and also within $N$-body simulations of the MW--LMC system \citep{Garavito-Camargo2019}.

\setlength{\tabcolsep}{7pt}
\renewcommand{\arraystretch}{1.5} 
\begin{table}
\centering
\caption{The reflex motion results using the H3+SEGUE+MagE outer halo survey data. These results are from the posterior conditioned on the mean and velocity dispersion data having removed discrepant data points, i.e., the red contours in all figures. We show the results evaluated using our linear continuity model at $r_{\rm gal} = [40, 100, 120] \, \rm kpc$. }
\begin{tabular}{lccc}
\hline
\hline
\input{tables/reflex-posterior}
\end{tabular}
\label{table:reflex-post}
\end{table}

\subsection{The velocity anisotropy of the stellar halo}

\begin{figure}
    \centering
    \includegraphics[width=\linewidth]{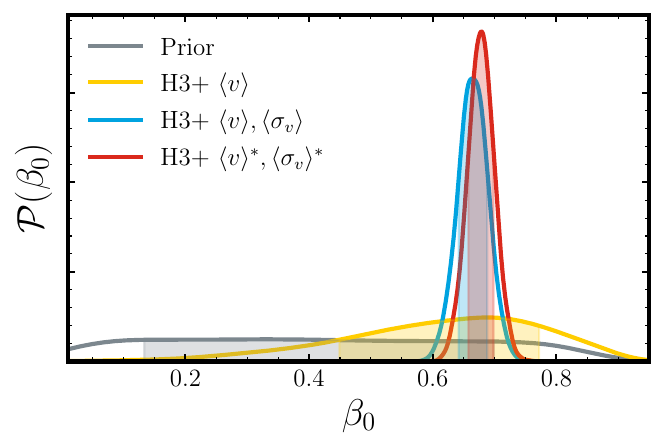}
    \caption{\textbf{Stellar halo velocity anisotropy:} The individual posteriors for the average velocity anisotropy of the stellar halo, $\beta_{0}$. 
    The yellow and blue contours represent the posteriors conditioned using the means, or means \textit{and} dispersions, of the stellar velocities, respectively.
    The red contours use the mean and dispersion data, but removing discrepant data points.
    The prior $1\sigma$ confidence interval is shown as the grey curve.}
    \label{fig7}
\end{figure}

The velocity anisotropy parameter, $\beta$, quantifies the radial bias of stellar velocities.
Previously, $\beta$ has been measured as a function of distance \citep{Cunningham2019, Bird2021, Han2024, Chandra2025a} largely suggesting that $\beta$ takes values $\gtrsim 0.6$ throughout the outer MW halo.
In Fig.~\ref{fig7} we show the posterior distributions for the average velocity anisotropy, $\beta_0$, conditioned on the mean only, or mean \textit{and} dispersion velocity data, as yellow and blue curves, respectively. 
The red curve uses the mean and dispersion data, but removing discrepant data points.
The shaded regions show the $16^{\rm th}-84^{\rm th}$ percentiles. 
The uniform simulation prior is shown in grey.
The expectation is that the velocity dispersion data is required to produce precise constraints on $\beta_0$. 
Indeed, this is what we see in Fig.~\ref{fig7} for the posteriors conditioned on dispersion data, returning a value $\beta_0 = 0.68 \pm 0.02$ after accounting for discrepant data.
This constraint is in agreement with previous measurements of the anisotropy \citep{Cunningham2019, Bird2021, Han2024}.
Interestingly, the posterior conditioned only on the mean velocity data also prefers a higher anisotropy. 
This slight preference can be understood \textbf{as} stellar haloes with different anisotropies naturally exhibiting different collective responses to a given perturbation
 e.g., via the interaction with the LMC \citep{Dillamore2026}.
Moreover, such a consistently large value for the velocity anisotropy throughout the MW halo, combined with the dynamical influence of the LMC, can align stars on highly eccentric orbits to explain known local structures in the MW \citep{Dillamore2026}.

\section{Discussion}\label{sec:discussion}

\subsection{The mass of the Milky Way and LMC in context}\label{sec:discussion-masses-in-context}

\subsubsection{The Milky Way}

\begin{figure}
    \centering
    \includegraphics[width=\linewidth]{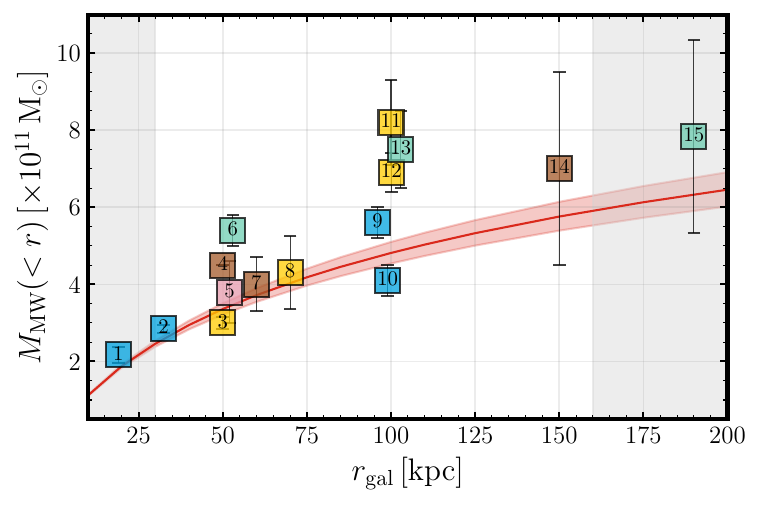}
    \caption{\textbf{Milky Way mass profile:} We show the posterior median and $1\sigma$ for the Milky Way enclosed mass as the red line and shaded region using the H3+SEGUE+MagE dataset. Regions outside of the radial range of the dataset, $r_{\rm gal} \notin [30 - 160] \, \rm kpc$, are shaded out. A selection of other previous literature values are coloured by the observational tracer used; blue for stellar streams, brown for BHB stars, pink for RRL stars, teal for satellites and globular cluster, and yellow for other stars. The numerical value indicates the referenced study:~
    1 -- \citet{Dillamore2025},
    2 -- \citet{Koposov2023}, 
    3 -- \citet{Williams2017},
    4 -- \citet{Williams2015},
    5 -- \citet{Medina2025}, 
    6 -- \citet{Vasiliev2019}, 
    7 -- \citet{Xue2008}, 
    8 -- \citet{Bird2022}, 
    9 -- \citet{Vasiliev2021},
    10 -- \citet{Gibbons2014}, 
    11 -- \citet{McMillan2017},
    12 -- \citet{Shen2022}, 
    13 -- \citet{CorreaMagnus2022},    
    14 -- \citet{Deason2012}, 
    15 --  \citet{Wang2022}. 
    Literature values with the same radial position are slightly separated for readability.}
    \label{fig8}
\end{figure}

To put the results from this study in context we compare with previous literature values using a range of techniques and dynamical tracers to constrain the enclosed MW mass. 
In Fig.~\ref{fig8} we show the MW mass enclosed posterior conditioned on the H3+SEGUE+MagE after removing discrepant data bins. 
The posterior median and $1\sigma$ are represented by the solid red lines and surrounding shaded area.
Each literature value is shown as a numbered square, see figure caption for references, with the colour representing the observational tracer used; blue for stellar streams, brown for Blue Horizontal Branch (BHB) stars, pink for RR Lyrae (RRL) stars, teal for satellites and globular cluster, and yellow for other stars.
For the enclosed mass at $50 \, \rm kpc$, as shown in more detail in our Fig.~\ref{fig3}, we find that our result is consistent with many others within uncertainty \citep[e.g.,][]{Williams2017, Koposov2023, Medina2025}. 
One of the most precise mass enclosed measurements at this distance is obtained by modelling the Orphan-Chenab stream in a combined MW--LMC potential, leading to an enclosed Milky Way mass, $M_{\rm MW}(<50\,\rm kpc) \approx 3.5 \, \pm \, 0.2 \times 10^{11} \, \rm M_{\odot}$ \citep{Koposov2023}. 
The fact that we also find a value consistent within $1\sigma$, and with a similar precision, is intriguing given the modelling scenarios are completely unique.
At $100 \, \rm kpc$ there is a large spread in previously reported values, although our posterior tends to agree closely with many at $\sim 5-6 \times 10^{11}\,\rm M_{\odot}$ \citep[e.g.,][]{Vasiliev2019, Vasiliev2021}.
The outermost data points are all consistent with our posterior within uncertainties, although we note our posterior is much more precise. 
However, the H3+SEGUE+MagE sample only covers distances out to $160 \, \rm kpc$, hence the posterior beyond this distance range is an extrapolation. 
Potentially, we are subject to overconfidence in our MW mass profile posterior due to the restrictive nature of the model parametrisation i.e., an NFW profile defined by 2 parameters $M_{200, \rm MW}$ and $ c_{200}$.

\subsubsection{The LMC}

\begin{figure}
    \centering
    \includegraphics[width=\linewidth]{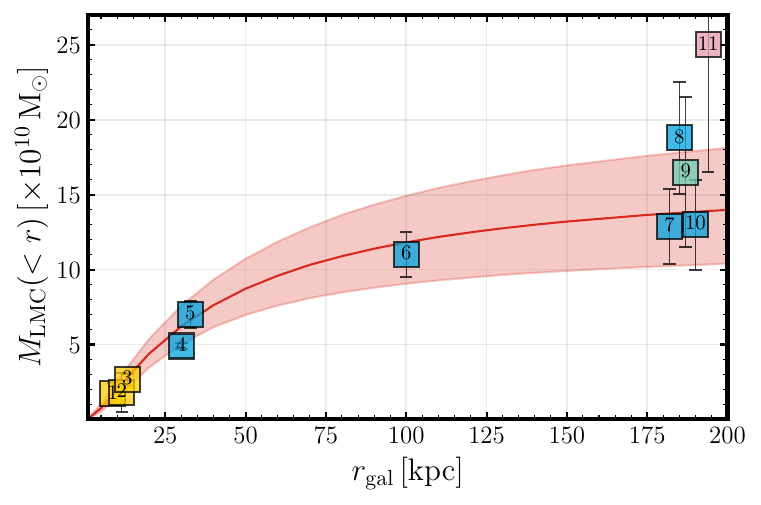}
    \caption{\textbf{LMC mass profile:} We show the posterior median and $1\sigma$ for the LMC enclosed mass as the red line and shaded region. Other previous literature values are shown as squares, coloured by the observational tracer used; blue for stellar streams, pink for Local Group, teal for satellites and yellow for LMC stars. The numerical value indicates the referenced study:~
    1 -- \citet{vanderMarel2014}, 
    2 -- \citet{Cullinane2020}, 
    3 -- \citet{Watkins2024}, 
    4 -- \citet{Warren2025}, 
    5/6/7 -- \citet{Koposov2023}, 
    8 -- \citet{Shipp2021}, 
    9 -- \citet{CorreaMagnus2022}, 
    10 -- \citet{Vasiliev2021}
    and 11 -- \citet{Penarrubia2016}.
    The outermost literature values are total mass estimates, separated on the $x$-axis for visualisation purposes.}
    \label{fig9}
\end{figure}

Similarly, we place our LMC mass results in context with other previous literature values using a range of techniques and dynamical tracers.
In Fig.~\ref{fig9} we show the LMC mass enclosed posterior conditioned on the H3+SEGUE+MagE after removing for discrepant data bins. 
The posterior median and $1\sigma$ are represented by the solid red lines and surrounding shaded area.
Each literature value is shown as a numbered square, see figure caption for references, with the colour representing the observational tracer used; blue for stellar streams, pink for Local Group, teal for satellites and yellow for LMC stars. 
Within $50 \, \rm kpc$, we find that our posterior mass profile prefers slightly more massive values but remains consistent with other studies within uncertainties \citep[e.g.,][]{Koposov2023, Warren2025}.
The outermost data points shown from previous studies all relate to total mass constraints for the LMC. 
We find that the total mass enclosed by our posterior agrees with other previous studies quoting $M_{\rm LMC} \sim  15\times 10^{10} \, \rm M_{\odot}$ 

\subsection{Caveats and Limitations}\label{sec:discussion-caveats}

Our approach of using rigid models for the MW--LMC systems comes with various caveats. Some of the limitations of rigid models include:~assuming that the MW is in equilibrium prior to the LMC's infall \citep[e.g.,][]{Deason2017}, assuming there is no mass lost or gained by either halo, not accounting for halo triaxiality, and using a simple Chandrasekhar prescription for dynamical friction \citep[see discussion sections in][]{Brooks2026, Brooks2025}.
Ideally, one would use a suite of $N$-body MW--LMC simulations that can naturally account for most of the above limitations. 
Moreover, $N$-body simulations would allow the MW disc to move separately from its dark matter halo, and account for deformation to the potentials, better modelling the induced reflex motion from the LMC \citep{Yaaqib2025}.
However, as we have shown, even state-of-the-art $N$-body simulations are in tension with the measured reflex motion parameters; see Fig.~\ref{fig5}.
In future studies, using a suite of $N$-body simulations \citep[e.g.,][]{Sheng2025} to train an SBI architecture to estimate the posterior distribution of e.g., the reflex motion parameters, will highlight any biases in the inference of these parameters when using rigid models to train similar SBI networks.

One of the other clear limitations of our rigid MW--LMC model set-up is the inability to explore large regions of simulation parameter space in the Galactocentric $y$-direction of the reflex motion, $v_y$. 
Despite accounting for radial variation in the reflex motion, even the allowed simulation prior space at a reference value of $100 \, \rm kpc$ is in tension with direct measurements from observational datasets \citep{Yaaqib2024, Bystrom2025, Chandra2025a}. 
One possible explanation is that the observational datasets contain undetected substructures which can produce larger values for $v_y$. 
For example, \citet{Chandra2025a} suggest that there may still be some LMC contamination in their data sample; see $3^{\rm rd}$ bin of the Q4 data in the middle panel of Fig.~\ref{fig1}.
Currently, our MW stellar haloes are comprised purely of MW star particles and do not contain any substructures. 
A future interesting avenue to pursue is to investigate the level of contamination required from, e.g., LMC stars, to produce simulated data points and values for $v_y$ which are consistent with the current H3+SEGUE+MagE dataset. 
Aside from undetected substructures, a rudimentary analysis of our rigid simulations shows that producing sufficiently large $v_y$ values is predominantly sensitive to LMC properties and the strength of dynamical friction.

SBI relies on the fact that any simulations used to train the inference include a sufficient amount of physics to properly describe the observations. 
The Predictive Posterior Check (PPC) carried out in this work, see Appendix.~\ref{sec:posterior-checks} and Fig.~\ref{figA2} \&~\ref{figA3}, show that the simulated data from our rigid MW--LMC models capture most of the observed data points well, although clearly some are not well represented. 
In Sec.~\ref{sec:results}, we removed any discrepant data points, those lying $2\sigma$ away from the simulation data mean values, and trained a separate SBI posterior estimator. 
For the returned MW mass, removing the discrepant velocity dispersion data points is important as it biased the mass estimate low by $\sim 10\%$.
It remains the scope of a future study to determine whether our rigid models are the limitation here, or whether there are persistent substructures remaining in the data to produce the observed values. 
Nevertheless, accounting for more of the observational selection effects e.g., better matching the exact spatial distribution of the observed data, and adopting a better set of summary statistics \citep[e.g., using spherical harmonic expansions for the velocity fields,][]{Cunningham2020} will help to create more realistic MW--LMC models used to perform the inference. 

Finally, the choice of prior distribution for model parameters can influence the final inference result. 
In \citet{Brooks2026} we found that using either an informative or uninformative prior on the MW and LMC total masses had a negligible effect on the inference. 
To remove the effect of confirmation bias on our posteriors, we used uninformative priors on the MW and LMC masses in this work.

\subsection{The influence of the LMC on Milky Way mass inference}\label{sec:discussion-lmc-importance}

In this section we investigate the importance of including an LMC within simulation models when estimating the MW mass using velocity dispersion data. 
Often, the MW is assumed to be in a state of dynamical equilibrium when using velocity dispersion measurements to constrain its mass \citep[e.g., Jeans modelling,][]{Jeans1915, Binney2008, CorreaMagnus2022, Medina2025}.
Using the initial distribution of stellar halo particles, one can calculate the radial and tangential velocity dispersion profiles \textit{prior} to the LMC's infall.
Therefore, we can measure the radial and tangential velocity dispersions using the same set of distance bins as before when computing the velocity dispersions from the MW stellar haloes evolved in the combined MW--LMC potential to present day; i.e., \textit{post} the LMC's infall.
Now, we are able to compare the bias introduced into the MW mass inference by using the velocity dispersion as measured either \textit{prior} or \textit{post} the LMC infall. 
This is equivalent to having \textit{excluded} or \textit{included} the LMC within our simulations.

To determine the level of bias in the inference, we train a separate neural network to estimate the MW mass using the H3+SEGUE+MagE velocity dispersion data, see Sec.~\ref{sec:data}, when adopting the velocity dispersions as those \textit{prior} the LMC infall. 
We adopt the same outer halo slope as in the inference, $\gamma = 4.55$ \citep{Amarante2024}.
All other simulation parameters remain unchanged.
When \textit{excluding} the LMC in the simulations, we find that the MW mass enclosed with $50\,\rm kpc$ prefers to be more massive by $\sim 5\% $ compared to the MW mass estimated when the LMC is \textit{included} in the simulations. 
Furthermore, the virial mass, $M_{200, \rm MW}$, is found to be overestimated by $\sim 10\% $  when the LMC is neglected in the simulations.
Therefore, the \textit{exclusion} of an LMC leads to the MW mass being biased to \textit{more} massive models.
This degree of bias is comparable to the measured precision of the MW mass in this work; see Sec.~\ref{sec:results}. 
Interestingly, our findings are similar to \citet{Erkal2020} who found that the MW mass is biased to larger values by $\sim 5 \%$ for $M_{\rm MW}(<50\,\rm kpc)$ when accounting for the mean velocity motions of stars i.e., the reflex motion, and averaging over stars distributed across the entire sky. Additionally, this level of bias is similar to that found by \citet{CorreaMagnus2022} using satellite tracer populations.
Therefore, within our presented methodology, including the LMC to account for dynamical disequilibrium is essential to model the reflex motion in the mean kinematics of MW halo stars; however, the velocity dispersions remain largely insensitive, yielding a relatively unbiased MW mass inference. 
This is similar to \citet{Sheng2025} who advocate that the LMC should be explicitly included in dynamical models, as mean stellar velocities primarily constrain the LMC, while the velocity dispersions constrain MW properties.

Furthermore, since velocity dispersions are proportional to the total mass of a system, the theoretical expectation is that dispersions prior to the LMC's infall will be smaller in magnitude than those computed post-infall.
To investigate this, we measured the dispersions prior to the LMC's infall for all of our simulations using the spherical Jeans equations.
We find that the dispersions are $\sim 5-10 \, \rm km \, s^{-1}$ smaller on average prior to the LMC infall. 
Interestingly, we note that these differences are prominently seen in the latitudinal component of the tangential velocity, $v_{t,b}$, and in the southern quadrants radial velocities, i.e., $v_{\rm GSR}$ in Q3 and Q4, whereas the northern quadrants, Q1 and Q2, remain largely unaffected.
This highlights the locality of the LMC's perturbative nature on the velocity dispersions of outer halo stars. 
Moreover, this is in agreement with \citet{Erkal2020} who also demonstrated that the stellar velocities in the northern hemisphere remain largely unaffected, allowing unbiased estimates of the MW mass when assuming equilibrium models.

\section{Conclusions \& Outlook}\label{sec:conclusions-outlook}

We have investigated the effect of the LMC in the MW via its perturbation to the kinematics of outer halo stars.
From the combined H3+SEGUE+MagE survey, we use the means and dispersions of the radial and tangential velocities of stars out to a distance of $160 \, \rm kpc$ to constrain properties of the MW, the LMC and their interaction.
We use an SBI framework to train neural posterior estimators using $32,000$ rigid MW--LMC simulations.
These MW--LMC simulations allow for disequilibrium to develop in the MW's stellar halo via their interaction.
Our SBI framework is able to directly map between observable quantities of the MW's stellar halo and simulation parameters.
After removing discrepant data relative to the simulated models, we find the following results and conclusions:

\begin{enumerate}
    \item  The enclosed mass of the MW and LMC within $50\,\rm kpc$ of their respective dynamical centres are determined to be:
    
        \begin{description}
            \item[\textbf{The enclosed MW mass:} $M_{\rm MW}(< 50 \, \rm kpc) = 3.36^{+0.15}_{-0.15} \times 10^{11}\, \rm M_{\odot}$] 
            \item[\textbf{The enclosed LMC mass:} $M_{\rm LMC}(< 50 \, \rm kpc) = 8.76^{+1.94}_{-1.77} \times 10^{10}\, \rm M_{\odot}$] 
        \end{description}
        \noindent 
        These new constraints are consistent within uncertainties of other previously reported values for both galaxies. 
    \vspace{0.4cm}
      \item The inference of the total mass of each halo implies a ratio of the total LMC infall mass to the total present-day MW mass of $\sim 30 \pm 10 \%$.
      \vspace{0.4cm}
      \item  The average velocity anisotropy of the stellar halo, prior to the LMC's infall is well constrained using the velocity dispersions of outer halo stars to be $\beta_{0} = 0.68 \pm 0.02$. This suggests that the MW's stellar halo was already radially anisotropic out to large distances prior to the LMC's recent passage. Moreover, this is consistent with, but generally lower than, the velocity anisotropy measured at present day in the Solar neighbourhood.
      \vspace{0.4cm}
      \item Using a linear continuity model to capture the radial variation in the reflex motion, we find the magnitude of the reflex motion, evaluated at a reference distance of $100\,\rm kpc$, to be:
      
        \begin{description}
            \item[\textbf{The reflex motion velocity:} $v_{\rm{travel}} = 38.6^{+8.3}_{-7.8} \, \rm km \, \rm s^{-1}$] 
        \end{description}
        \noindent 
        Meanwhile, the direction of the reflex motion remains challenging to constrain, further highlighting the inconsistencies with and between other previous measurements. 

    \vspace{0.4cm}
    \item If the influence of the LMC is not accounted for in the simulations, we find that the MW virial mass is biased to larger values by up to $\sim10 \%$. Hence, while the LMC is crucial for modelling dynamical disequilibrium in the mean halo kinematics, it has minimal impact on velocity dispersion-based constraints of the MW mass.

 \end{enumerate}

Our analysis has produced some of the most precise results to date on properties of the MW and LMC, as well as their relative interaction via the reflex motion.
In the near future, there will be an increase in the raw number of outer halo stars from SDSS-V \citep[][]{SDSSCollaboration2025}. Moreover, this SDSS-V catalogue can be combined with distance and proper motion measurements, which have much improved precision, from Gaia DR4.
Both of these factors, increased precision and number counts of outer halo stars, will improve the inference performed using this SBI framework, provided that observational facilities remain the dominant source of uncertainty.

\section*{Acknowledgements}

We thank the reviewer for a detailed and constructive report that improved this study.
RANB acknowledges support from the Royal Society and would like to thank Dennis Zaritsky and Ben Johnson for comments on the manuscript as well as the entire H3 team for access to their data.
JLS acknowledges the support of the Royal Society (URF\textbackslash R1\textbackslash191555; URF\textbackslash R\textbackslash 241030). 
NGC acknowledges support from the Heising-Simons Foundation grant \#2022-3927, through which NGC is supported by the Barbara Pichardo Future Faculty Fellowship.

\section*{Data Availability}

The MW--LMC simulation parameters and the data for $32,000$ unique stellar haloes, each with $20,000$ particles, will be shared upon reasonable request. 
A Jupyter notebook to make the stereographic projection of the reflex motion apex direction can be found \href{https://gist.github.com/dc-broo3/f246dc4824b43d6a20b8a122f9d29a92}{here}.

\noindent
\textit{Software:} \texttt{sbi} \citep{tejero-cantero2020sbi}, \texttt{agama} \citep{2019MNRAS.482.1525V}, \texttt{gala} \citep{gala}, NumPy \citep{harris2020array}, Matplotlib \citep{Hunter:2007}, Seaborn \citep{Waskom2021}, Astropy \citep{astropy:2013, astropy:2018, astropy:2022}, SciPy \citep{2020SciPy-NMeth}.

\bibliographystyle{mnras}
\bibliography{biblio} 


\appendix

\section{Posterior Diagnostic Checks}\label{sec:posterior-checks}

Any posterior from a generative approach should be assessed for its accuracy through a variety of diagnostic tests to gain trust that the inference has been successfully performed.
To trust the results presented in this work we carry out posterior coverage probability checks and predictive posterior checks.

\subsection{Coverage probability test}

A coverage probability test is one way to assess the accuracy of an estimated posterior.
In Bayesian analysis, a coverage test checks whether credible intervals have the expected probabilities \citep[see,][sec.~2.4 for a concise explanation]{Jeffrey2025}. 
In a 1-dimensional posterior setting, one can define a particular credible interval to be
the narrowest interval containing, for example, $68\%$ of the probability weight.
The Bayesian inference procedure takes in some observed data, $D_{\rm obs}$, and determines a posterior distribution, $p(\theta | D_{\rm obs})$, and hence a credible interval for $\theta$. 
For a coverage test, one uses a test parameter, $\theta_{\rm test}$, selected from the prior, $p(\theta)$, as the input to a simulation that produces the corresponding output data point, $D_{\rm test}$. From this, one can derive a posterior, $p(\theta | D_{\rm test})$, and therefore a credible interval. If the inference process has been correctly implemented, then the true test parameter value, $\theta_{\rm test}$, will fall in this credible interval, in this example, $68\%$ of the time. 
Repeating this test for many sampled $\theta_{\rm test}$, and varying the credibility intervals, one can gain trust that the estimated posterior distributions are accurate and have reliable confidence intervals.

To perform a coverage test on our SBI posteriors, we use ‘Tests of Accuracy with Random Points' \citep[TARP,][see their figs.~1\&2 for further intuition]{Lemos2023} as implemented in the \texttt{sbi} Python package \citep{tejero-cantero2020sbi}.
For our application of SBI, this test is relatively straightforward as we have many pre-existing simulations with an amortised inference scheme, i.e., each data evaluation is computationally cheap without the need to retrain the neural network \citep{Mittal2025}.
TARP coverage probabilities test the accuracy of estimated posteriors by only using samples from the posterior. 
This technique is similar to simulation based calibration \citep{Talts2018} but extends the idea to the full-dimensional posterior space instead of being restricted to 1-dimension. 

\begin{figure}
    \centering
    \includegraphics[width=\linewidth]{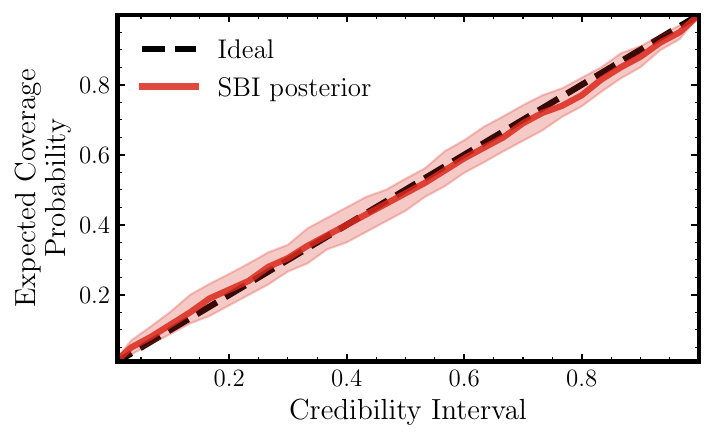}
    \caption{\textbf{Coverage probability posterior check:} For the SBI posterior conditioned on the mean and dispersion H3+SEGUE+MagE data, after removing discrepant data points, we show the probability of finding true test parameters in the appropriate credible intervals matches the expected coverage probability. A bootstrapped $1\sigma$ uncertainty is shown as the shaded region.
    This validates the estimated SBI posterior distributions and allows one to interpret the confidence intervals on parameter constraints as representative and reliable.}
    \label{figA1}
\end{figure}

In Fig.~\ref{figA1}, we demonstrate that the expected coverage does indeed match the credibility level for the estimated posteriors in this work conditioned using both the mean \textit{and} dispersion velocity data from H3+SEGUE+MagE.
This validates our neural posterior estimation as being truly representative of the probability that each of our model parameters has some true value with truly representative uncertainties.
This can be further quantified in two ways. 
Firstly, we can compute the area between the ideal TARP curve and our posterior TARP curves for credibility intervals greater than $0.5$; namely, the Area To Curve (ATC) value. This number should be close to $0$, a value $\gg0$ indicates an estimated posterior that is too wide, conversely, a value $\ll0$ indicates that the estimated posterior is too narrow. 
Secondly, we can calculate the p-value of a Kolmogorov-Smirnov test. The null hypothesis is that an exact one-to-one curve and our posterior TARP curve are identical. If this p-value is less than $0.05$, then this null hypothesis is rejected. 
For our SBI posterior we report an ATC magnitude $\lesssim 0.1$ and a Kolmogorov-Smirnov p-value of $1.0$. 
These values suggest that we are not drastically over-/under-fitting, or biased, and are not required to reject the posterior. 

\subsection{Predictive posterior check}

We carry out a ‘Posterior Predictive Check' (PPC) to act as a complementary diagnostic test. This test makes use of the fact that if the inference has been correctly implemented, then any generated data, $D_{\rm pp}$, using simulation parameters as sampled from the posterior, $\theta_{\rm pp}$, should be similar to the observed data, $D_{\rm obs}$ \citep[][]{Lueckmann21a}. 
A PPC provides intuition about any bias introduced in inference. For example, it can help to determine whether or not the generated data systematically differ from the observed data used during the inference.
We carry out a PPC for the posterior conditioned on the mean \textit{and} dispersion velocity data from H3+SEGUE+MagE.
To do this, we first sample model parameter values from these posterior distributions. 
We then re-run rigid MW--LMC simulations adopting these parameter values, generating output data as the mean and dispersion of the radial and tangential velocity measurements, which we can then use to compare to the observed data that was originally used to condition the posterior. 

\begin{figure}
    \centering
    \includegraphics[width=\linewidth]{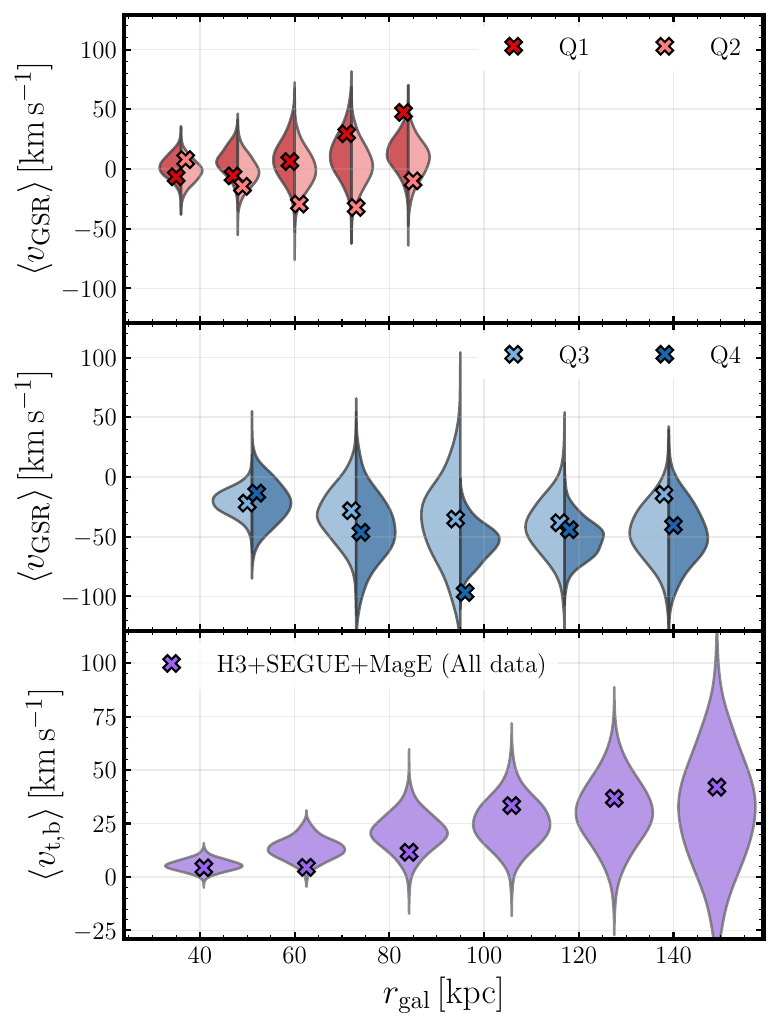}
    \caption{\textbf{Predictive Posterior Check - mean velocities:}
    We test the quality of the SBI posterior returned using the H3+SEGUE+MagE data.
    We show the mean radial (top and middle panels for northern and southern quadrants, respectively) and tangential (bottom panel) velocity data as generated using samples from this posterior as the violin plot contours. 
    The observed data points are shown as the same coloured crosses.
    We do not show the data errorbars as the generated data already accounts for the survey uncertainties.
    The generated and original data look sufficiently similar implying the SBI posteriors are representative of the observed data.}
    \label{figA2}
\end{figure}

\begin{figure}
    \centering
    \includegraphics[width=\linewidth]{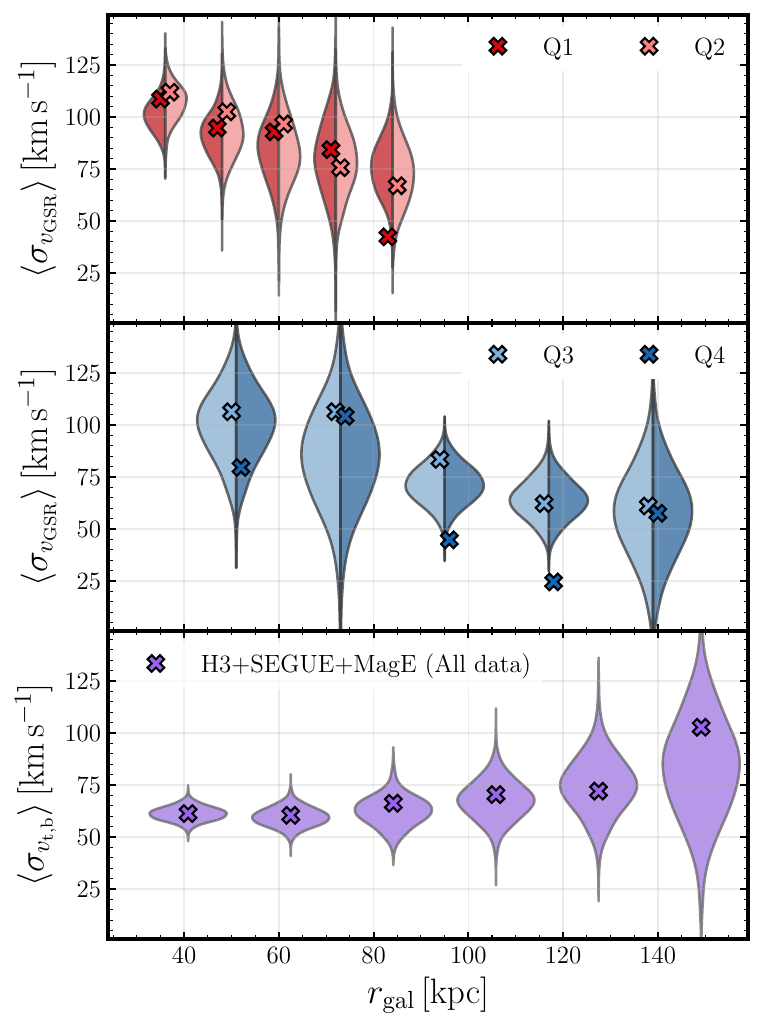}
    \caption{\textbf{Predictive Posterior Check - velocity dispersions:}
    We test the quality of the SBI posterior returned using the H3+SEGUE+MagE data.
    We show the dispersion of the radial (top and middle panels for northern and southern quadrants, respectively) and tangential (bottom panel) velocity data as generated using samples from this posterior as the violin plot contours. 
    The observed data points are shown as the same coloured crosses.
    We do not show the data errorbars as the generated data already accounts for the survey uncertainties.}
    \label{figA3}
\end{figure}

In Fig.~\ref{figA2} and Fig.~\ref{figA3}, we demonstrate the PCC for the posteriors conditioned on the full H3+SEGUE+MagE dataset i.e.,. open blue contours in all figures.
Using parameters sampled from this posterior, we show the generated data for the radial and tangential, mean and dispersion, velocity data as the violin plot contours.
The mean observed data points are shown as the same coloured markers; same points as in Fig.~\ref{fig1} and Fig.~\ref{fig2}.
Measurement errors are not shown because the generated data incorporates survey uncertainties.
In most distance bins the observed data is found to be well represented by the generated simulation data.
However, there are clearly some data points which are difficult to reconcile with the simulations. For example, the outermost bins in Q1 for the mean radial velocity, or similarly for the 3rd/4th Q4 bins for the radial velocity dispersions, are hard to explain with the current modelling.  
This is likely a limitation of using reasonably simple rigid MW--LMC simulations which are able to capture global velocity perturbations, but struggle to capture smaller scale perturbations.
Although, even $N$-body deforming simulations \citep{Garavito-Camargo2019}, struggle to reproduce these velocity trends in the northern Galactic hemisphere \citep[see fig.~4,][]{Chandra2025a}.
Potentially, some of these discrepancies between the observed and generated data could be due to unresolved substructure across these distances, for example stars stripped from the LMC \citep{Chandra2025a}. 

One of the advantages of SBI is the flexibility of choosing which data to train the inference network on. To investigate any biases introduced by including all data points in our inference, we removed data points that lie $2\sigma$ away from the mean of each simulation distribution.
With this reduced set of velocity summary statistics for outer halo stars, we re-train the inference framework, resulting in the red posterior distributions shown throughout all figures in Sec.~\ref{sec:results}; see that section for more details on the degree of any bias introduced. 

\bsp	
\label{lastpage}
\end{document}

%% file: tables/priors.tex
& \\
\(\log_{10}( M _{200,\rm MW} )\) & \(  \mathcal{U}(10.5, 12.5) \, M_{\odot} \) \\
\(  M_{\rm LMC} \) & \( \mathcal{U}(2.5, 50) \times 10^{10} M_{\odot} \) \\
& \\
\( \log_{10}(\lambda_{\mathrm{DF}}) \) & \( \mathcal{U}(-3, 1) \) \\
& \\
\( \alpha_{\rm LMC} \) & \(\mathcal{U} (72.5^\circ, 87.5^\circ) \) \\
\( \delta_{\rm LMC} \) & \(\mathcal{U}(-75^\circ, -65^\circ)  \) \\ 
\( d_{\rm LMC} \) & \( \mathcal{N}(49.6, 2)\,\mathrm{kpc}\) \\
\( v_{\rm los} \) & \( \mathcal{N}(262.2, 5) \,\mathrm{km}\,\mathrm{s}^{-1} \) \\
\( \mu_{\alpha_{\rm LMC}} \) & \( \mathcal{N}(1.9, 0.25)\, \mathrm{mas}\,\mathrm{yr}^{-1} \) \\ 
\( \mu_{\delta_{\rm LMC}} \) & \( \mathcal{N}(0.33, 0.25)\,\mathrm{mas}\,\mathrm{yr}^{-1} \) \\
& \\
\(\beta_0 \) & \(  \mathcal{U}(0, 0.85)\) \\
\( r_{\rm dens} \) & \(\mathcal{U}(10, 20)\, \rm kpc \) \\
\( \gamma\) & \(\mathcal{U}(3, 5) \) \\
\hline

%% file: tables/reflex-posterior.tex

Parameter & $r_{\rm gal} = 40 \, \rm kpc$ & $r_{\rm gal} = 100 \, \rm kpc$ & $r_{\rm gal} = 120 \, \rm kpc$ \\
\hline
\vspace{-0.5cm}\\
$v_{\rm travel} \, [\rm km \, \rm s^{-1}]$ & $17.0^{+21.1}_{-7.0}$ & $38.6^{+8.3}_{-7.8}$ & $47.2^{+11.7}_{-11.0}$  \\
$l_{\rm apex} \, [ ^{\circ}]$ & $2.5^{+79.5}_{-71.0}$ & $-20.0^{+58.4}_{-51.0}$ & $-17.1^{+65.0}_{-57.7}$  \\
$b_{\rm apex} \, [ ^{\circ}]$ & $-60.4^{+32.3}_{-17.5}$ & $-75.9^{+20.5}_{-7.0}$ & $-76.4^{+21.3}_{-7.4}$  \\
\vspace{-0.3cm}\\
\hline
\vspace{-0.3cm}\\
$v_{x} \, [\rm km \, \rm s^{-1}]$ & $2.8^{+10.3}_{-4.7}$ & $5.3^{+6.7}_{-5.1}$ & $5.8^{+8.1}_{-6.4}$  \\
$v_{y} \, [\rm km \, \rm s^{-1}]$ & $0.1^{+10.5}_{-6.8}$ & $-2.2^{+7.6}_{-6.8}$ & $-1.9^{+8.8}_{-8.8}$  \\
$v_{z} \, [\rm km \, \rm s^{-1}]$ & $-12.9^{+6.9}_{-13.8}$ & $-36.0^{+9.4}_{-7.7}$ & $-44.1^{+13.3}_{-10.8}$  \\
\vspace{-0.3cm}\\
\hline